\documentclass{elsart}
\usepackage{amsmath,amssymb,graphicx}
\journal{Nuclear Physics A}

\newcommand{\Ng}{N_{\text{g}}}
\newcommand{\Ngg}{N_{\text{gg}}}
\newcommand{\Nc}{N_{\text{c}}}
\newcommand{\Qs}{Q_{\text{s}}}

\newcommand{\rmi}{\mathrm{i}}
\newcommand{\rmd}{\mathrm{d}}
\newcommand{\rme}{\mathrm{e}}

\newcommand{\ktilde}{\widetilde{\boldsymbol{k}}_\perp}
\newcommand{\kbar}{\overline{\boldsymbol{k}}_\perp}
\newcommand{\phat}{\hat{\boldsymbol{p}}_\perp}

\newcommand{\zerot}{\boldsymbol{0}_\perp}
\newcommand{\p}{\boldsymbol{p}}
\newcommand{\pt}{\boldsymbol{p}_\perp}
\newcommand{\q}{\boldsymbol{q}}
\newcommand{\qt}{\boldsymbol{q}_\perp}
\newcommand{\kt}{\boldsymbol{k}_\perp}
\newcommand{\ktone}{\boldsymbol{k}_{1\perp}}
\newcommand{\kttwo}{\boldsymbol{k}_{2\perp}}
\newcommand{\xt}{\boldsymbol{x}_\perp}
\newcommand{\yt}{\boldsymbol{y}_\perp}
\newcommand{\zt}{\boldsymbol{z}_\perp}
\newcommand{\del}{\boldsymbol{\partial}_\perp}
\newcommand{\Xt}{\boldsymbol{X}_\perp}

\newcommand{\bb}{\boldsymbol{b}}
\newcommand{\bx}{\boldsymbol{x}}

\newcommand{\dely}{\Delta y}

\newcommand{\Ac}{\mathcal{A}}

\newcommand{\G}{\mathcal{G}}
\newcommand{\Gpm}{\mathcal{G}_{(>)}}
\newcommand{\Gmp}{\mathcal{G}_{(<)}}

\newcommand{\rhoA}{\rho_{\text{A}}}
\newcommand{\rhop}{\rho_{\text{p}}}
\newcommand{\freeR}{\Delta_{\text{R}}}

\newcommand{\tr}{\mathrm{tr}}

\newcommand{\tf}{T_{\text{F}}}
\newcommand{\ta}{T_{\text{A}}}
\newcommand{\dpt}[1]{\frac{\mathrm{d}^2\boldsymbol{p}_{#1\perp}}{(2\pi)^2}}
\newcommand{\dqt}[1]{\frac{\mathrm{d}^2\boldsymbol{q}_{#1\perp}}{(2\pi)^2}}
\newcommand{\dk}[1]{\frac{\mathrm{d}^2\boldsymbol{k}_{#1\perp}}{(2\pi)^2}}
\newcommand{\dprimek}[1]{\frac{\mathrm{d}^2\boldsymbol{k}'_{#1\perp}}{(2\pi)^2}}

\begin{document}

\vspace*{-5mm}
\begin{flushright}
{\scriptsize YITP-08-47, RBRC-741}
\end{flushright}
\vspace{5mm}

\begin{frontmatter}
\title{Two-gluon Production and Longitudinal Correlations
       in the Color Glass Condensate}
\author{Kenji Fukushima}
\address{Yukawa Institute for Theoretical Physics,
         Kyoto University, Oiwake-cho, Kitashirakawa,
         Sakyo-ku, Kyoto 606-8502, Japan}
\author{Yoshimasa Hidaka}
\address{RIKEN BNL Research Center,
         Brookhaven National Laboratory,\\
         Upton, New York 11973, USA}
\begin{abstract}
 We derive an analytical expression for the two-gluon production in
 the pA (light-heavy) collisions, and focus specifically on the
 rapidity dependent part.  We approximate the gauge field from the
 heavy target as the Color Glass Condensate which interacts with the
 light projectile whose source density allows for a perturbative
 expansion.  We discuss the longitudinal correlations of produced
 particles.  Our calculation goes in part beyond the eikonal limit for
 the emitted gluons so that we can retain the exponential terms with
 respect to the rapidity difference.  Our expression can thus describe
 the short-range correlations as well as the long-range ones for which
 our formula is reduced to the known expression.  In a special case of
 two high-$p_t$ gluons in the back-to-back kinematics we find that
 dependence on the rapidity separation is only moderate even in the
 diagrammatically connected part.
\end{abstract}
\end{frontmatter}


\section{Introduction}

  The gluon distribution function in hadrons grows up at high
collision energy and thus small Bjorken's $x$, which makes it
indispensable to elaborate a resummation over the multiple scattering
to abundant soft gluons.  When the transverse distribution of partons
is saturated eventually at small $x$ and small
$Q^2$~\cite{saturation}, the classical approximation is to be a good
description of the hadron wave-function dominated by coherent
small-$x$ partons.  In fact, a shift in the perturbative QCD (Quantum
Chromodynamics) vacuum by such a classical field is an efficient
prescription to take account of dense gluons within the eikonal
approximation.  This resummation scheme has been well founded in the
non-linear but still perturbative regime embodied by the notion of the
Color Glass Condensate (CGC)~\cite{McLerran:1993ni,review}.

  In this paper we are interested in the inclusive one- and
two-gluon production cross section in the CGC formalism.  It is not
quite viable to solve the non-linear problem fully in the case that
the target and the projectile both have dense gluon contents.  In
contrast, the analytical calculation is feasible if we deal with the
asymmetric case of a light projectile scattering off a heavy
target~\cite{review:pA}, as is the case in the pA collision or in the
forward and backward rapidity regions in the AA collision.  We remark
that the inclusive multi-particle production has been discussed in
several different frameworks~\cite{Leonidov:1999nc,Kharzeev:2003sk,%
Tuchin:2004rb,JalilianMarian:2004da,Blaizot:2004wu,Baier:2005dv,%
Kovchegov:2006qn,Fujii:2006ab,Li:2007zzc,Marquet:2007vb,Braun:2008zs,%
Gelis:2008rw}.  The present work aims to elucidate the production rate
of two gluons having two distinct rapidities $y_1$ and $y_2$ in the
asymmetric collisions.  We have to limit the separation $|y_1-y_2|$
parametrically less than $1/\alpha_{\text{s}}$ since we do not take
account of quantum evolution which is incorporated by the radiative
corrections between two separate gluons.  It is maybe worth noting
that, nevertheless, we can describe another aspect of quantum
evolution through the hadron wave-function, that is, the evolution of
the saturation scale~\cite{Iancu:2002aq}, though it is non-trivial
whether and how these quantum corrections are to be
factorized~\cite{Gelis:2008rw}.

  We shall accomplish the whole calculation of the two-gluon
production along the line of
Refs.~\cite{Blaizot:2004wu,Gelis:2005pt,Blaizot:2008yb} and thus our
results should be an extension of that given in
Ref.~\cite{JalilianMarian:2004da} where $y_1\gg y_2$ is assumed for
simplicity.  Our results are in fact consistent with
Ref.~\cite{Baier:2005dv} in the eikonal limit that emitted gluons are
energetic enough to overwhelm the transverse recoil, while they
possess intricate extra contributions depending on $\dely=y_1-y_2$
with the eikonal limit relaxed.  We will later discuss what the
missing part was that causes the discrepancy.

  It should be an important problem to clarify the particle
correlation reflecting the parton saturation effect.  This is
because the CGC configuration naturally leads to strong
chromo-electric and chromo-magnetic fields which extend between the
target and the projectile~\cite{Kovner:1995ja,Lappi:2006fp}.  It would
be conceivable that the longitudinal correlation provides us with
information on the initial dense state to be regarded as CGC matter.

  Moreover, in Ref.~\cite{Kharzeev:2004bw}, the jet azimuthal
back-to-back correlation has been considered, which is also discussed
in Refs.~\cite{Baier:2005dv,Marquet:2007vb}.  Also, from different
interest, the recent experimental data at RHIC (Relativistic Heavy-Ion
Collider) exhibits the wide rapidity correlation in the near side of
the trigger and associated
jets~\cite{Putschke:2007mi,Adams:2005aw,Daugherity:2006hz,Wenger:2008ts},
which could have an influential remnant from the initial state and the
strong color field in
it~\cite{Majumder:2006wi,Dumitru:2008wn,Gavin:2008ev}.  In a slightly
different context, in Refs.~\cite{Armesto:2006bv,Armesto:2007ia}, the
forward-backward correlation has been discussed which seems to have a
direct link to the longitudinal field presumably existing in CGC matter.  
Such a long-range rapidity correlations have been found in
the recent experiment at RHIC~\cite{Srivastava:2007ei,Li:2008gn}.  All
these preceding efforts are decent to motivate us to carry out the
analytical QCD evaluation in the same procedure as
Ref.~\cite{Blaizot:2004wu} where only minimal assumptions under
theoretical control hide behind modeling.

  In the subsequent sections we shall elucidate the two-gluon
correlation starting with the established computation for the
inclusive one-gluon cross section for self-contained discussions.


\section{Gluon Production}

  We here make a quick review on the formulation to evaluate the gluon
production amplitude, which should be useful for clarity of our
convention adopted in the present paper.


\subsection{One-gluon production}

  The production amplitude of one on-shell gluon with three-momentum
$\p$, helicity $\lambda$, and color $a$ is expressed as the amplitude
from the initial state $|\Omega_\text{in}\rangle$ to the one-gluon
asymptotic state $|\p,\lambda,a\rangle$.  Using the
Lehmann-Symanzik-Zimmermann (LSZ) reduction formula~\cite{Peskin}, we
can generally write this amplitude in a form of
\begin{equation}
 \langle \p;\lambda,a|\Omega_\text{in}\rangle = \rmi\,
  \epsilon^{(\lambda)}_\mu(\p)\,\int\rmd^4x\,
  \rme^{\rmi p\cdot x}\,\square_x\,
  \langle\Omega_\text{out}|A^\mu_a(x)|\Omega_\text{in}\rangle \,,
\label{eq:single_amp}
\end{equation}
where $A^\mu_a(x)$ denotes the gluon field operator.  We note that we
should take the on-shell limit $p_0\to E_p=|\p|$ in the right-hand
side because the emitted gluon is real hence on-shell.  The phase
space integral and the summation over physical degrees of freedom lead
us to the following expression for the multiplicity of emitted gluons
per rapidity~\footnote{Precisely speaking this expression is not the
  multiplicity but the \textit{probability} for the one-particle
  emission.  Nevertheless, this simply coincides with the multiplicity
  in the leading order of $g$ in the present pA problem.  We do not
  distinguish the multiplicity and the one-particle emission
  probability here for this reason.  In the case of the AA collision,
  however, all multiple-scattering processes contribute to the
  multiplicity~\cite{Gelis:2006yv}.};
\begin{equation}
 \left\langle \frac{\rmd\Ng}{\rmd y} \right\rangle
  = \frac{1}{4\pi}\int\dpt{} \sum_{\lambda,a}
  \bigl\langle \bigl|\langle\p;\lambda,a|\Omega_\text{in}
  \rangle \bigr|^2 \bigr\rangle \,.
\label{eq:dndy}
\end{equation}
Here we have introduced the momentum rapidity variable as
$y\equiv\half\ln[p^+/p^-]$ where $p^\pm\equiv(p^0\pm p^z)/\sqrt{2}$
which can be expressed as $|\pt|\,\rme^{\pm y}/\sqrt{2}$ by using the
on-shell condition, i.e.\ $2p^+p^-\!=|\pt|^2$.  It is then
straightforward to confirm that the phase space integral is rewritten
as $\rmd^3\p/[(2\pi)^3 2E_p]=\rmd y\rmd^2\pt/[2(2\pi)^3]$.  In
Eq.~(\ref{eq:dndy}) the summation with respect to $\lambda$ should run
over only the transverse components corresponding to physical degrees
of freedom, and $\langle\cdots\rangle$ denotes an average over the color
configuration of sources.

  It is convenient to adopt the following convention to define the
physical polarization in the light-cone
coordinates~\cite{Srivastava:2000cf}.  Let us particularly consider
the light-cone gauge defined by
\begin{equation}
 n_\mu A^\mu = A^+ = 0 \,,
\label{eq:gauge}
\end{equation}
where $n_\mu\equiv \delta^+_{\;\mu}$.  Then, as beautifully demonstrated in
Refs.~\cite{Gelis:2005pt,Hatta:2006ci}, we can take advantage of this
gauge choice to reduce computational cost significantly.  The
light-cone gauge condition leads to $\epsilon^{(\lambda)+}=0$ and
there are not four but three independent polarization vectors, two of
whose linear combinations are physical.  The following projection
operator enables us to identify the transverse parts simply with
$\lambda=1,2$, that is,
\begin{equation}
 \epsilon^{(\lambda)}_\mu (p) \equiv -D_{\mu\lambda}(p) \,,
\label{eq:transverse}
\end{equation}
where
\begin{equation}
 D_{\mu\nu}(p)\equiv -g_{\mu\nu} + \frac{n_\mu p_\nu+n_\nu p_\mu}
  {n\cdot p} - \frac{p^2}{(n\cdot p)^2}n_\mu n_\nu \,.
\label{eq:D}
\end{equation}
The above tensor structure appears also in the gluon propagator in the
light-cone gauge.  It is important to note that these polarization
vectors are doubly transverse to both the momentum and gauge
directions, i.e.\
\begin{equation}
 p^\mu \epsilon_\mu^{(\lambda)}(p) = 0 \,,\qquad
 n^\mu \epsilon_\mu^{(\lambda)}(p) = 0 \,.
\label{eq:double_trans}
\end{equation}
Moreover, using the explicit form of Eqs.~(\ref{eq:transverse}) and
(\ref{eq:D}), we can show that the sum over physical polarization
amounts to
\begin{equation}
 \sum_{\lambda=1,2} \epsilon^{(\lambda)}_\mu(p)
  \epsilon^{(\lambda)}_\nu(p) = D_{\mu\nu}(p) \,.
\label{eq:physical_sum}
\end{equation}
Since the above is multiplied by $A^\mu$ and $A^\nu$ after all (see
Eq.~(\ref{eq:single_amp})), only a quantity of $A^\mu D_{\mu\nu}A^\nu$
is necessary to reach the final expression.  This quantity is reduced
to as simple as $A^i A^i$ owing to the gauge
condition~(\ref{eq:gauge}).  Therefore, we only have to retain a
$\delta_{ij}$ term out of $D_{\mu\nu}(p)$ in the polarization
summation to estimate the one-gluon production.  Consequently the
gluon multiplicity is expressed
as~\cite{Blaizot:2004wu,Gelis:2005pt,Kovchegov:1998bi,%
Dumitru:2001ux,Kovchegov:2001sc,Kharzeev:2003wz}
\begin{equation}
 \left\langle \frac{\rmd\Ng}{\rmd y} \right\rangle
 = \frac{1}{4\pi}\int\dpt{}\; (p^2)^2\left\langle\Ac_a^{i\ast}(p)
  \,\Ac_a^i(p) \right\rangle\,,
\label{eq:single}
\end{equation}
where we have denoted the vacuum expectation value of the gauge field,
i.e.\ background gauge field, as
$\Ac^\mu_a(x)\equiv\langle\Omega|A^\mu_a(x)|\Omega\rangle$.  Because
$p^2$ in Eq.~(\ref{eq:single}) is a four-momentum squared, only the
singular contribution proportional to $1/p^2$ in $\Ac^i_a(p)$
remains non-vanishing in the on-shell limit that we should take in the
end.  We note that we define the Fourier transformation in our
convention as
\begin{equation}
 \Ac_a^i(p) \equiv \int\rmd^4x\,\rme^{\rmi p\cdot x}
  \Ac_a^i(x) \,.
\end{equation}
Generally this is a complex-valued function and satisfies
$\Ac_a^{i\ast}(p)=\Ac_a^i(-p)$.


\subsection{Two-gluon production}

  It is a straightforward generalization to continue on considering
not only single but also the multiple gluon production in this way.
We will focus particularly on the two-gluon case in this work, though
it is in principle possible to accommodate more gluons.  In the same
way as previously explained, we can write down the production
amplitude for two gluons with three-momenta $\p$, $\q$, helicity
$\lambda$, $\sigma$, and color $a$, $b$, respectively.  The LSZ
reduction formula yields the amplitude as
\begin{equation}
 \begin{split}
 & \langle\p,\lambda,a;\q,\sigma,b|\Omega_\text{in}\rangle \\
 & = \rmi\,\epsilon^{(\lambda)}_\mu(\p)\int\rmd^4x\,
  \rme^{\rmi p\cdot x}\,\square_x\;
  \rmi\,\epsilon^{(\sigma)}_\nu(\q)\int\rmd^4y\,
  \rme^{\rmi q\cdot y}\,\square_y\;
  \langle\Omega_\text{out}|\mathrm{T}[A^\mu_a(x)A^\nu_b(y)]
  |\Omega_\text{in}\rangle \,.
 \end{split}
 \label{eq:LSZ}
\end{equation}
Here $\mathrm{T}$ represents the time ordering as usual.  From this
amplitude we can reach an expression for the two-gluon correlation in
the leading order of $g$ for one emitted with rapidity $y_1$ and the
other emitted with rapidity $y_2$;
\begin{align}
 &\left\langle\frac{\rmd\Ng}{\rmd y_1 \rmd y_2} \right\rangle
  = \frac{1}{16\pi^2}\int \dpt{}\dqt{}
  \sum_{\lambda,\sigma,a,b}\bigl\langle \bigl|
  \langle\p,\lambda,a;\q,\sigma,b|\Omega_\text{in}\rangle \bigr|^2
  \bigr\rangle \notag\\
 &= \left\langle\frac{\rmd\Ng}{\rmd y_1}\frac{\rmd\Ng}{\rmd y_2}
  \right\rangle  +\frac{1}{16\pi^2}\int \dpt{}\dqt{}\,(p^2)^2\,(q^2)^2\,
 \left\langle 2\text{Re}\bigl[\Ac^{\ast i}_a(p)\,\G_{ab}^{ij}(p,q)\,
  \Ac^{\ast j}_b(q) \bigr] \right\rangle\notag\\
 &\quad +\frac{1}{16\pi^2}\int \dpt{}\dqt{}\,
  (p^2)^2 \, (q^2)^2 \,\left\langle \G_{ba}^{\dagger\,ji}(q,p)\, 
  \G_{ab}^{ij}(p,q)  \right\rangle  \,.
\label{eq:dndydy}
\end{align}
The first term corresponds to a long-range correlation part which is
not suppressed for large $|y_1-y_2|$.  This term is important to
consider the azimuthal angle correlation, which arises through the
average over the CGC configurations~\cite{Dumitru:2008wn}.  The second
is an interference term, and the third is a short-range correlation
part which depends on $\dely$ as diminishing as $|\dely|$ goes large.
To extract the connected contribution henceforth, we have defined the
propagator and its Fourier transform as follows;
\begin{align}
 \G^{\mu\nu}_{ab}(x,y) &\equiv \langle\Omega_\text{out}|\mathrm{T}
  [A^\mu_a(x) A^\nu_b(y)]|\Omega_\text{in}\rangle
  -\Ac^\mu_a(x)\Ac^\nu_b(y) \,,\\
 \G^{\mu\nu}_{ab}(p,q) &\equiv \int\rmd^4x\,\rmd^4y\;
  \rme^{\rmi p\cdot x +\rmi q\cdot y}\; \G^{\mu\nu}_{ab}(x,y) \,.
\end{align}
Just for clarity we remark that $\G_{ba}^{\dagger\,ji}(q,p)$ literally
means $\G_{ab}^{\ast\,ij}(p,q)$.  The central discussion in what
follows lies in analytical evaluation of the background gauge field in
the CGC formalism and the associated propagator on top of the
background.


\begin{figure}
\begin{center}
 \includegraphics[width=11cm]{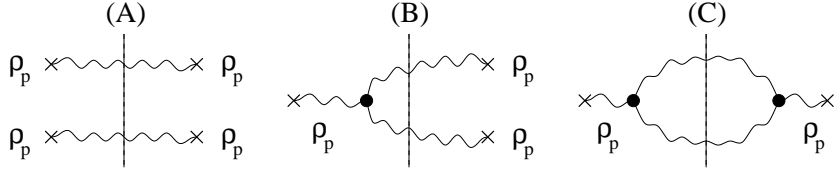}
 \caption{Diagrammatic representation of three distinct contributions
 to the two-gluon production: (A) the disconnected
 $\mathcal{O}(\rhop^4)$ diagram, (B) the interference
 $g\mathcal{O}(\rhop^3)$ one, and (C) the connected
 $g^2\mathcal{O}(\rhop^2)$ one.}
 \label{fig:production}
\end{center}
\end{figure}


  We can pictorially visualize these three terms enumerated in
Eq.~(\ref{eq:dndydy}) in a way as sketched in
Fig.~\ref{fig:production}.  The first (disconnected), the second
(interference), and the third (connected) terms of
Eq.~(\ref{eq:dndydy}) correspond to Figs.~\ref{fig:production}~(A),
(B), and (C), respectively.  The cross represents the interaction with
the light projectile whose intrinsic color distribution is regarded as
small perturbation, i.e.\ $\rhop=\mathcal{O}(g)$.  As a result of this
setting (A), (B), and (C) have the same order;
(A)$=\mathcal{O}(\rhop^4)$, (B)$=g\mathcal{O}(\rhop^3)$,
(C)$=g^2\mathcal{O}(\rhop^2)$, and they are all $\mathcal{O}(g^4)$.
We can drop (B) under an assumption of the Gaussian average over
$\rhop$.  Although the long-range correlation stemming from (A), which
is predominant in the AA collision, is physically
interesting~\cite{Dumitru:2008wn,Armesto:2006bv,Gelis:2006yv}, we will
not consider (A) and concentrate in (C) only that is the contribution
carrying the short-range correlations like $\rme^{\pm\dely}$.  In fact,
what we will pursue in this work is to retrieve the neglected rapidity
dependence in the two-gluon production with emphasis on the analytical
formulation.  We will report the full correlation including (A) and
phenomenological implication in a separate
publication~\cite{hidaka:2008}.

  We shall comment that the leading-order piece with the rapidity
dependence is identified differently according to whether it is the pA
or the AA collision.  The rapidity dependence in the pA case comes
from the connected diagram in the leading order, while the AA
collision has no rapidity dependence because the leading-order diagram
is not the connected but the boost-invariant disconnected one.  Hence,
the rapidity dependence in the AA collision starts from the
next-leading order where the logarithmic terms in rapidity appear.  It
is possible to resum all the leading-logarithmic terms only to
renormalize the weight function of the color source distribution,
resulting in the evolution by means of the JIMWLK
equation~\cite{Gelis:2008rw}.  The same treatment for the
leading-logarithmic terms could be applicable to the disconnected
diagram in the pA case.


\section{Background Gauge Field with Dense-Dilute Sources}

  We shall adopt the CGC picture to proceed to concrete calculations.
The background field originates from the dense color distribution
inside a fast-moving hadron.  In a classical approximation in which
quantum corrections (i.e.\ the wave-function renormalization to the
field expectation value) are neglected, $\Ac^\mu(x)$ is simply a
solution to the classical Yang-Mills equations of motion with color
source $\rhoA(\xt)$.  This type of the one-source problem is solvable
and the explicit form of $\Ac^\mu(x)$ as a function of $\rhoA(\xt)$ is
known~\cite{Kovchegov:1996ty}.

  In this paper, as we mentioned in Introduction, we shall consider
a situation with two color sources corresponding to asymmetric
collisions;  one source distribution is dense from a heavy nuclei
which gives the color current $J^-=\delta(x^+)\rhoA(\xt)$ and the
other is dilute from a proton for instance, which gives
$J^+=\delta(x^-)\rhop(\xt)$.  We treat $\rhop(\xt)$ as small
perturbation, where we should note that the classical approximation
still works because of the presence of large $\rhoA(\xt)$.

  Here, these naive color currents do not satisfy the covariant
conservation.  We thus need to augment these definitions with
appropriate gauge rotations as
\begin{equation}
 J_a^- = \delta(x^+) V_{ab}(x){\rhoA}_b(\xt) \,,\qquad
 J_a^+ = \delta(x^-) U_{ab}(x){\rhop}_b(\xt) \,,
\label{eq:current}
\end{equation}
where $V(x)=\mathcal{P}_{x^-}\exp[\,\rmi g\int^{x^-}\rmd z^- A^+]$ and
$U(x)=\mathcal{P}_{x^+}\exp[\,\rmi g \int^{x^+}\rmd z^+ A^-]$ with
$\mathcal{P}_{x^\pm}$ being the path-ordering with respect to $x^\pm$.
The color matrices $A^\pm$ are the full gauge fields (including both
classical and quantum parts) in the color adjoint representation.  The
covariant conservation is then manifest because $D^+ V(x)=0$ and
$D^- U(x)=0$.

  The analytical calculation makes sense when either $\rhoA(\xt)$ or
$\rhop(\xt)$ is small compared to the gluon transverse momentum
(i.e.\ saturation scale).  We can then perform the perturbative
expansion in terms of small $\rhop(\xt)$.  The gauge field associated
with such an asymmetric collision can have an expansion like
\begin{equation}
 \Ac^\mu = \Ac_{(0)}^\mu + \Ac_{(1)}^\mu + \cdots \,,
\label{eq:decomposition}
\end{equation}
where $\Ac_{(0)}^\mu$ and $\Ac_{(1)}^\mu$ are the zeroth and first
order terms with respect to $\rhop(\xt)$, respectively.  One can in
principle continue the expansion up to arbitrary order, but what we
need to retain in this work is only up to the first order.  In this
formalism of the pA problem the gauge fields become time independent
after interaction at the collision point, as we will see shortly.  
The LSZ formula means that the residue of the massless pole
corresponds to the probability of the gluon associated with the gauge
fields in the asymptotic state.  This makes a contrast to the
approximate prescription often used in the AA case;  the field
amplitude for each momentum mode translates into the particle number,
which could depend on time~\cite{Lappi:2003bi}.  It is not easy how
the LSZ formula works for the AA case~\cite{Gelis:2006yv}, in
particular in the expanding geometry, while there is no ambiguity in
the pA collision.


\subsection{Gauge choice}

  It is practically crucial to take an appropriate gauge choice to
make the whole computational procedure transparent.  In what follows,
we will make use of the light-cone (LC) gauge $A^+=0$ with the
background field $\Ac_{(0)}^\mu$ which is usually referred to as a
solution in the covariant (COV) gauge,
\begin{equation}
 \Ac_{(0)}^+=0\,,\quad
 \Ac_{(0)}^-=-\frac{1}{\del^2}\delta(x^+)\rhoA(\xt)\,,\quad
 \Ac_{(0)}^i = 0\,.
\label{eq:background}
\end{equation}
It should be mentioned that $V(x)=1$ in this $A^+=0$ gauge so that no
complication enters involving $\rhoA(\xt)$, which is a great advantage
in this technique.  In contrast to that, even in the leading order in
$\rhop(\xt)$ at the classical level, we have to keep $U(x)$ which
stems from nonzero $\Ac_{(0)}^-$.  Because $\Ac_{(0)}^-$ has
$\delta(x^+)$ in it, the $x^+$ dependence of $U(x)$ is given as
\begin{equation}
 U(x) = \left\{ \begin{array}{p{1mm}l}
  & 1 \quad \text{for } x^+<0 \,,\\
  & \displaystyle
  U(\xt)\equiv \mathcal{P}_{x^+}\exp\biggl[\,\rmi g
  \int_{0^-}^{0^+}\!\rmd x^+ \Ac_{(0)}^-\biggr]
  \quad \text{for } x^+\ge0^+ \,.
 \end{array} \right.
\end{equation} 
We will use the same notation for $U(x)$ and $U(\xt)$ hoping that no
confusion would arise from this.  We remark that the integration with
respect to $x^+$ in the exponential needs a careful treatment with
appropriate
regularization~\cite{Blaizot:2004wu,Fukushima:2007dy,Fukushima:2007ki}.

  To reiterate explicitly, our choice (that was first adopted in
Ref.~\cite{Gelis:2005pt}) is
\begin{equation}
 \Ac^\mu = \overbrace{\underbrace{\Ac_{(0)}^\mu}_{\text{COV for A}}
  + \Ac_{(1)}^\mu}^{\text{LC for p}} .
\label{eq:decomposition2}
\end{equation}
The essential point in this trick invented in Ref.~\cite{Gelis:2005pt}
is that the classical solution in the COV gauge for left-moving
$\rhoA(\xt)$ is consistent with the LC gauge for right-moving
$\rhop(\xt)$.  Actually, since $\Ac_{(0)}^+=0$, the
solution~(\ref{eq:background}) obtained in the COV gauge for
$\rhoA(\xt)$ may well be regarded as a solution in the LC gauge for
$\rhop(\xt)$.  [Usually $\Ac_{(0)}^+=0$ is \textit{assumed} in
addition to the gauge fixing condition.]

  The benefits from the choice (\ref{eq:decomposition2}) are twofold:
One is understood from Eq.~(\ref{eq:physical_sum}), that is, the
light-cone gauge significantly simplifies the summation over physical
degrees of freedom.  The other is the fact that only one component of
the background field~(\ref{eq:background}) is non-vanishing and the
remaining $\Ac_{(0)}^-$ takes a finite value at $x^+=0$ only.
Therefore the gluon propagation receives no effect from the background
field except at $x^+=0$ where the eikonal phase and color rotation are
induced.  This nice feature tremendously reduces the computational
labor.


\subsection{Classical solution}

  We are evaluating $\Ac_{(1)}^\mu$ that is necessary not only for the
one-gluon production but also for the two-gluon production.  
From the equations of motion, $D_\mu F^{\mu\nu}=J^\nu$,
in the light-cone gauge we have a set of equations to determine
$\Ac_{(1)}^\mu$, that is,
\begin{align}
 & -\partial^{+2}\Ac_{(1)}^- + \partial^+\partial^i \Ac_{(1)}^i
  = J^+ \,,
\label{eq:nu+}\\
 & \partial^+ D_{(0)}^- \Ac_{(1)}^- -\del^2 \Ac_{(1)}^-
  +D_{(0)}^-\partial^i \Ac_{(1)}^i -2\rmi g
  [\partial^i \Ac_{(0)}^-,\Ac_{(1)}^i] = 0 \,,
\label{eq:nu-}\\
 & 2\partial^+ D_{(0)}^- \Ac_{(1)}^i -\partial^+\partial^i
  \Ac_{(1)}^- -\del^2 \Ac_{(1)}^i +\partial^i\partial^j \Ac_{(1)}^j
  = 0 \,,
\label{eq:nui}
\end{align}
for $\nu=+,-,i$, respectively.  It is notable that $J^-$ disappears
in the second equation.  Using Eq.~(\ref{eq:nu+}) we can simplify
Eq.~(\ref{eq:nu-}) into
\begin{equation}
 \square \Ac_{(1)}^- -2\rmi
  g\bigl[\Ac_{(0)}^-,\partial^+ \Ac_{(1)}^-\bigr]
  -2\rmi g\bigl[\partial^i \Ac_{(0)}^-,\Ac_{(1)}^i\bigr] = 0 \,.
\label{eq:nu-_s}
\end{equation}
Here we used $D_{(0)}^-J^+=0$.  In the same way Eq.~(\ref{eq:nui})
becomes as simple as
\begin{equation}
 \square \Ac_{(1)}^i -2\rmi
  g\bigl[\Ac_{(0)}^-,\partial^+ \Ac_{(1)}^i\bigr]
  = -\frac{\partial^i}{\partial^+}J^+ \,.
\label{eq:nui_s}
\end{equation}
It is not difficult to solve Eq.~(\ref{eq:nui_s}) for $\Ac_{(1)}^i$
with the initial boundary condition,
\begin{equation}
 \Ac_{(1)}^i(x^+<0,x^-,\xt) = \theta(x^-)\frac{\partial^i}{\del^2}
  \rhop(\xt) \,,
\end{equation}
which is nothing but a retarded solution in the presence of one
right-moving source before the collision.

  Let us then examine the connection condition across the singularity
located at $x^+=0$.  By integrating Eq.~(\ref{eq:nui_s}) with respect
to $x^+$ from $0^-$ to $0^+$, and picking up a singularity contained in
$\Ac_{(0)}^-$ (see Eq.~(\ref{eq:background})), we can acquire the
boundary condition,
\begin{equation}
 \int_{0^-}^{0^+}\rmd x^+ \,D_{(0)}^- \Ac_{(1)}^i = 0 \,.
\end{equation}
Here we have dropped the overall $\partial^+$ which is irrelevant.  We
immediately see the solution to the above equation to be
\begin{equation}
 \Ac_{(1)}^i(x^+=0^+,x^-,\xt) = U(\xt)
  \Ac_{(1)}^i(x^+=0^-,x^-,\xt) \,,
\label{eq:bc}
\end{equation}
remembering that $D^-U(x)=0$.  In the forward light-cone region,
i.e.\ $x^+>0$, there is no background field and thus the gluon
propagation is expressed by the retarded free propagator $\freeR$
defined by $\square_x\freeR(x)=\rmi\delta^{(4)}(x)$ leading to
\begin{equation}
 \freeR(x) = \int\frac{\rmd^4 p}{(2\pi)^4}\,
  \frac{-\rmi}{p^2 + \rmi p^+\epsilon}\,\rme^{-\rmi p\cdot x} \,.
\end{equation}
Here $\epsilon\to0^+$ is understood as usual.  It is easy to make sure
that $\freeR$ is nonzero only for $x^+>0$.  It is straightforward to
confirm from the above definition a useful relation,
\begin{equation}
 \partial^+\freeR(x)\bigr|_{x^+=0^+}
  =\frac{\rmi}{2}\delta(x^-)\delta^{(2)}(\xt) \,.
\label{eq:propagator_del}
\end{equation}
Then the solution to Eq.~(\ref{eq:nui_s}) at nonzero $x^+$ is
generally to be expressed as
\begin{equation}
 \Ac_{(1)}^i =  \Ac_{(1<)}^i + \Ac_{(1>)}^i
\end{equation}
with
\begin{align}
 \Ac_{(1<)}^i(x) &= \theta(-x^+)\theta(x^-)\frac{\partial^i}{\del^2}
  \rhop(\xt) \,,\\
 \Ac_{(1>)}^i(x) &= \int_{y^+>0}\rmd^4y\,\bigl[-\rmi\,\freeR(x-y)
  \bigr]\biggl[-\frac{\partial^i_y}{\partial^+_y}J^+(y)\biggr]
  \notag\\
  & \qquad +\int \rmd y^-\rmd^2\yt\,\bigl[-2\rmi\partial^+_x
  \freeR(x-y)\bigr]\Ac_{(1)}^i(y)\Bigr|_{y^+=0^+} \,.
\label{eq:a_i}
\end{align}
Here $\Ac_{(1<)}^i$ is the past solution coming from the source for
$x^+<0$.  The first term in $\Ac_{(1>)}^i$ has the same origin but it
involves a color rotation by $U(\xt)$ additionally for $x^+>0$.  In
the presence of the second term $\Ac_{(1>)}^i$ satisfies the correct
boundary condition~(\ref{eq:bc}) in the $x^+\to0^+$ limit, which is
obvious from Eq.~(\ref{eq:propagator_del}).  Also we can readily find
that this second term produces no contribution for $x^+>0$ when acted
by the d'Alembertian $\square$.  From these expressions we can operate
the d'Alembertian to find
\begin{align}
 \square \Ac_{(1<)}^i &= -2\delta(x^+)\delta(x^-)
  \frac{\partial^i}{\del^2}\rhop(\xt)
  - \theta(-x^+)\theta(x^-)\partial^i\rhop(\xt) \,,\\
 \square \Ac_{(1>)}^i &= 2\delta(x^+)\delta(x^-)U(\xt)
  \frac{\partial^i}{\del^2}\rhop(\xt)
  - \theta(x^+)\theta(x^-)\partial^i\bigl[U(\xt)\rhop(\xt)\bigr] \,.
\end{align}
We can take the Fourier transform and reach
\begin{align}
 -p^2 \Ac_{(1<)}^i(p) &= -\rmi p^i
  \biggl[\frac{p^2}{(p^+ +\rmi\epsilon)(p^- -\rmi\epsilon)}\biggr]
  \frac{\rhop(\pt)}{\pt^2} \,, \\
 -p^2 \Ac_{(1>)}^i(p) &= -\rmi\int\dk{1} \biggl[\frac{p^i\ktone^2}
  {(p^+ +\rmi\epsilon)(p^- +\rmi\epsilon)} - 2k_1^i\biggr]
  U(\kttwo)\frac{\rhop(\ktone)}{\ktone^2} \,,
\label{eq:sol_ai}
\end{align}
where $\ktone+\kttwo=\pt$.  We can compute the gluon
multiplicity by substituting the sum of the above expressions into
Eq.~(\ref{eq:single}) and taking an ensemble average over the
$\rhop(\xt)$ and $\rhoA(\xt)$ distributions.  We will not do that
because the final results are
known~\cite{Blaizot:2004wu,Dumitru:2001ux} and the physical quantity
of our current interest is the two-gluon correlation.

  Although it is not relevant to the one-gluon case, the two-gluon
production requires the functional form of $\Ac_{(1)}^-$ for the
present purpose.  This is actually where our results come to differ
from Ref.~\cite{Baier:2005dv}.  There is no source contribution
$\propto J^\mu$ in $\Ac_{(1)}^-$.  From Eq.~(\ref{eq:nu-_s}) the
boundary condition is inferred from
\begin{equation}
 \int_{0^-}^{0^+}\rmd x^+ \,\Bigl( D_{(0)}^-\Ac_{(1)}^-
  -\rmi g\Bigl[\partial^i \Ac_{(0)}^-,
  \frac{1}{\partial^+}\Ac_{(1)}^i\Bigr]\Bigr) = 0 \,,
\end{equation}
where we have dropped the overall $\partial^+$ in the same way as
previously.  The above leads to the connection condition as
\begin{equation}
 \begin{split}
 \Ac_{(1)}^-(x^+=0^+,x^-,\xt) &=
  U(\xt)\Ac_{(1)}^-(x^+=0^-,x^-,\xt)\\
 &\qquad + \bigl[\partial^i U(\xt)\bigr]
  \frac{1}{\partial^+}\Ac_{(1)}^i(x^+=0^-,x^-,\xt) \,.
 \end{split}
\label{eq:bc2}
\end{equation}
The first term in the right-hand side is zero in fact because
$\Ac_{(1)}^-(x^+=0^-)=0$.  Nevertheless, we write it to avoid
confusion in order to define a convenient notation $M^\mu_{\;\nu}$ in
later discussions.  The solution for arbitrary $x^+$ is, therefore,
given as
\begin{align}
 \Ac_{(1<)}^-(x) &= 0 \,,\\
 \Ac_{(1>)}^-(x) &= \int \rmd y^-\rmd^2\yt \bigl[-2\rmi
  \partial_x^+\freeR(x-y)\bigr] \Ac_{(1)}^-(y) \Bigr|_{y^+=0^+} \,.
\end{align}
The representation in momentum space is thus $\Ac_{(1<)}^-(p)=0$ and
\begin{equation}
 -p^2 \Ac_{(1>)}^-(p) = -\rmi\int\dk{1}\,\frac{-2\ktone\cdot\kttwo}
  {p^+ +\rmi\epsilon}\,U(\kttwo)\frac{\rhop(\ktone)}{\ktone^2} \,,
\label{eq:sol_a-}
\end{equation}
where $\ktone+\kttwo=\pt$ again.  We note that the right-hand side has
no dependence on $p^-$ because $\square \Ac_{(1>)}^-$ is proportional
to $\delta(x^+)$.  For consistency check one can readily see that
$\Ac_{(1>)}^i$ and $\Ac_{(1>)}^-$ obtained above certainly satisfy the
rest of the equations of motion, namely, Eq.~(\ref{eq:nu+}) as they
should.

  Later on, for notational simplicity, we will introduce $M^\mu_{\;\nu}$
and $C^\mu$ to express the connection condition and the solution,
respectively.


\section{Propagator with the Background Gauge Field}

  It is the background propagator that is necessary for the two-gluon
correlation.  We will calculate it here by means of the expansion in
terms of small $\Ac_{(1)}^\mu \propto\rhop$.  The full background
propagator without expansion is defined by
\begin{equation}
 -\rmi \bigl[ D^\lambda D_\lambda \,g_{\mu\nu}
  -D_\mu D_\nu -2\rmi g F_{\mu\nu} \bigr]_x
  \G^{\nu\sigma}(x,y) = g_\mu^{\;\sigma} \delta^{(4)}(x-y) \,,
\label{eq:def_prop}
\end{equation}
where both $\Ac_{(0)}^-$ and $\Ac_{(1)}^\mu$ enter the background
field.  We will use the adjoint matrix notation from now on, namely, a
color matrix is understood as
\begin{equation}
 (\Ac^\mu)_{ab} = (\ta^c \Ac^\mu_c)_{ab}
  \equiv \rmi f^{acb}\Ac^\mu_c \,,
\end{equation}
in the color adjoint representation.  Then, the quantity in the square
brackets is decomposed into the zeroth-order and first-order parts
with respect to $\Ac_{(1)}^\mu$.  It follows that we have
\begin{equation}
 \begin{split}
 & D^\lambda D_\lambda g_{\mu\nu} - D_\mu D_\nu -2\rmi g F_{\mu\nu} \\
 &\qquad = D_{(0)}^\lambda D_{(0)\lambda} g_{\mu\nu}
   -D_{(0)\mu}D_{(0)\nu} - 2\rmi g F_{(0)\mu\nu}
   + \rmi \delta\Gamma_{\mu\nu} +\mathcal{O}(\rhop^2)\,,
 \end{split}
\label{eq:decompose}
\end{equation}
where we have defined the vertex matrix of order
$\mathcal{O}(\rhop)$ as follows;
\begin{equation}
 \rmi\delta\Gamma_{\mu\nu} \equiv
  -\rmi g\Bigl\{ \bigl[ (\partial_\lambda \Ac_{(1)}^\lambda)
  +2\Ac_{(1)}^\lambda\partial_\lambda \bigr]g_{\mu\nu}
  + 2(\partial_\mu \Ac_{(1)\nu})
  - 2(\partial_\nu \Ac_{(1)\mu}) \Bigr\} \,.
\label{eq:Gamma}
\end{equation}
We remark that we have dropped terms irrelevant to the gluon
propagator, that means, terms which vanish when sandwiched by the
zeroth-order propagator with transversality and also with the upper
$+$ components vanishing due to gauge fixing.  We shall define the
zeroth-order background propagator as
\begin{equation}
 -\rmi \bigl[D_{(0)}^\lambda D_{(0)\lambda}\, g_{\mu\nu}
  -D_{(0)\mu} D_{(0)\nu} -2\rmi g F_{(0)\mu\nu} \bigr]_x
  \G^{\nu\sigma}_{(0)}(x,y) = g_\mu^{\;\sigma}\delta^{(4)}(x-y) \,,
\end{equation}
and then we can express the full propagator in Eq.~(\ref{eq:def_prop})
in a way as $\G=(\G_{(0)}^{-1}+\delta\Gamma)^{-1}$, that yields an
expansion,
\begin{equation}
 \G^{\mu\nu}(x,y) = \G^{\mu\nu}_{(0)}(x,y)
  -\int d^4z\,\G^{\mu\lambda}_{(0)}(x,z)\,
  \delta\Gamma_{\lambda\sigma}(z)\,\G^{\sigma\nu}_{(0)}(z,y)
  +\mathcal{O}(\rhop^2)\,.
\label{eq:prop}
\end{equation}
At this stage it is tangible how we dropped terms from
Eq.~(\ref{eq:decompose}) to Eq.~(\ref{eq:Gamma});  irrelevant terms in
$\delta\Gamma_{\mu\nu}$, in fact, arise from the transverse
properties, i.e., $n_\mu\G_{(0)}^{\mu\nu}=0$ and
$\partial_\mu\G_{(0)}^{\mu\nu}=0$ (see also
Eq.~(\ref{eq:double_trans})).  In momentum space we can write the
vertex as
\begin{equation}
 \delta\Gamma_{\mu\nu}(p,q) = \rmi g \Ac^\lambda_{(1)}(p+q)
 \Gamma_{\mu\lambda\nu}(p,q)
\label{eq:vertex}
\end{equation}
with
\begin{equation}
 \Gamma_{\mu\lambda\nu}(p,q) = (p-q)_\lambda\, g_{\mu\nu}
  +2q_\mu g_{\nu\lambda} - 2p_\nu g_{\lambda\mu} \,.
\end{equation}
It should be mentioned that the above vertex is slightly different
from a standard QCD three-point vertex.  It might seem that the
momentum conservation does not hold.  This is because $q_\nu$ out of
$-2p_\nu-q_\nu$ vanishes from $q_\nu\G_{(0)}^{\nu\rho}(q)=0$ and
$p_\mu$ out of $2q_\mu+p_\mu$ vanishes from
$\G_{(0)}^{\rho\mu}(p)p_\mu=0$.  All the terms involving $\Ac_{(0)}^-$
disappear due to gauge fixing.  We shall then concretely compute
Eq.~(\ref{eq:prop}) in the subsequent subsections below.


\begin{figure}
\begin{center}
\includegraphics[width=11cm]{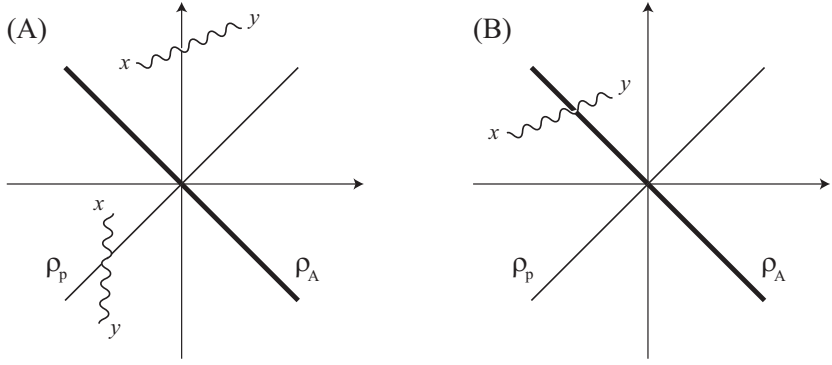}
\end{center}
\caption{Some typical examples for $\G_{(0)}^{\mu\nu}(x,y)$:  (A) The
  propagation is free from the source singularity in the case for
  ($x^+<0$, $y^+<0$) or ($x^+>0$, $y^+>0$).  (B) The singularity at
  $x^+=0$ affects the propagation which occurs in the case for
  ($x^+<0$, $y^+>0$) or ($x^+>0$, $y^+<0$).}
\label{fig:G0}
\end{figure}



\subsection{$\G_{(0)}^{\mu\nu}(x,y)$ for $(x^+>0,\;y^+>0)$ or
  $(x^+<0,\;y^+<0)$}

  Now let us calculate the propagator in the region without crossing
the source singularity, that is, the region with $x^+>0$ and $y^+>0$
or the region with $x^+<0$ and $y^+<0$ as depicted in
Fig.~\ref{fig:G0}~(A).  There, the background propagator is just the
free one which is unity in color space given by
\begin{equation}
 -\rmi \bigl[(\partial^\lambda \partial_\lambda)g_{\mu\nu}
  -(\partial_\mu \partial_\nu) \bigr]_x \Delta^{\nu\sigma}(x,y)
  =g_\mu^{\;\sigma}\delta^{(4)}(x-y) \,.
\end{equation}
In the momentum representation, we can write the time-ordered
propagator in a concise form as
\begin{equation}
 \Delta^{\mu\nu}(p,q) \equiv (2\pi)^4 \delta^{(4)}(p+q)
  \frac{\rmi D^{\mu\nu}(p)}{p^2+\rmi \epsilon} \,,
\end{equation}
using $D^{\mu\nu}(p)$ defined in Eq.~(\ref{eq:D}).  The standard
$\epsilon$ prescription enables us to define the retarded
($x^+\!>\!y^+$) and advanced ($x^+\!<\!y^+$) propagators, i.e.,
$\Delta_{\text{R}}^{\mu\nu}(p,q)$ and
$\Delta_{\text{A}}^{\mu\nu}(p,q)$ by replacement of the denominator by
$p^2+\rmi\epsilon p^+$ and $p^2-\rmi\epsilon p^+$, respectively.


\subsection{$\G_{(0)}^{\mu\nu}(x,y)$ for $(x^+<0<y^+)$ or
  $(y^+<0<x^+)$}

  As a preparation to proceed to the case with a singularity at
$x^+=0$ on the propagating path, we shall introduce the following
notation.  We will express the boundary conditions, Eq.~(\ref{eq:bc})
and Eq.~(\ref{eq:bc2}), into a single form of
\begin{equation}
 A^\mu(x^+=0^+) = M^\mu_{\;\;\nu} A^\nu(x^+=0^-) \,,
\label{eq:bc_gauge}
\end{equation}
where the gauge rotation associated with the singularity is
\begin{equation}
 \begin{split}
 & M^i_{\;j} = \delta^i_{\;j} U(\xt)\,,\quad
  M^-_{\;-} = U(\xt)\,,\quad
  M^-_{\;i} = \partial^i U(\xt)\frac{1}{\partial^+}\,,\\
 & M^i_{\;-} = M^i_{\;+} = 0\,,\quad
  M^-_{\;+} = 0\,,\quad
  M^+_{\;\mu} = 0\,.
 \end{split}
\end{equation}
This representation is useful in the following formulation.

  If $x^+<0<y^+$ or $y^+<0<x^+$ then the propagator passes over the
source singularity.  We show an example in Fig.~\ref{fig:G0}~(B).  We
will derive the appropriate boundary condition at the singularity
here.  To do so, we shall pick up only $\partial^-$ and $\delta(x^+)$
terms from the equations of motion.  Then we consider the connection
condition for each polarization component $\mu$ for the case of
$y^+<0<x^+$ first.  From the boundary condition for the gauge
field~(\ref{eq:bc_gauge}) it is straightforward to write the boundary
condition down for the propagator as well in a similar form as
\begin{equation}
 \G_{(0)}^{\mu\nu}(x,y)\Bigr|_{x^+=0^+}
  = M^\mu_{\;\sigma}\,\Delta^{\sigma\nu}(x,y) \Bigr|_{x^+=0^-} \,.
\end{equation}
We note that the free propagator appears in the right-hand side for
$x^+=0^-$ and $y^+<0$.  Using the same technique as in the previous
section where we calculated the gauge field, we can express the
propagator for $y^+<0<x^+$ with satisfying the above boundary
condition as
\begin{align}
 \Gpm^{\mu\nu}(x,y) &= \int\rmd z^-\rmd^2 \zt\,\bigl[ -2\rmi
  \partial_x^+ \freeR^{\mu\lambda}(x,z)\bigr]\,g_{\lambda\sigma}\,
  \G_{(0)}^{\sigma\nu}(z,y)\Bigr|_{z^+=0^+} \notag\\
 &= -\int\rmd z^-\rmd^2\zt\,\bigl[-2\rmi\partial_x^+
  \freeR^{\mu i}(x,z)\bigr] U(\zt)\,\Delta^{i\nu}(z,y)\Bigr|_{z^+=0}
  \,.
\label{eq:prop_r}
\end{align}
In the momentum representation the above translates into
\begin{align}
 \Gpm^{\mu\nu}(p,q) &= \int\rmd^4x\rmd^4y\,
  \rme^{\rmi p\cdot x + \rmi q\cdot y}\,
  \Gpm^{\mu\nu}(x,y)\, \theta(-y^+) \notag\\
 &=(2\pi)\delta(p^+ \!+ q^+)\,\theta(p^+)
  \frac{2p^+ D^{\mu\lambda}(p) D_\lambda^{\;\nu}(q)}
  {(p^2+\rmi\epsilon p^+)(q^2-\rmi\epsilon q^+)}U(\pt\!+\qt) \,.
\label{eq:proppm}
\end{align}
Here we note that the propagator poles are located only on the lower
(or upper) half plane in terms of the complex $p^-$ (or $q^-$
respectively) variable, which reflects the boundary condition,
$y^+<0<x^+$.

Similarly, the propagator for $x^+<0<y^+$ is to be written as
\begin{equation}
 \Gmp^{\mu\nu}(p,q)
  =(2\pi)\delta(p^+ \!+ q^+)\,\theta(q^+)
  \frac{2q^+ D^{\mu\lambda}(p) D_\lambda^{\;\nu}(q)}
  {(p^2 -\rmi\epsilon p^+)(q^2 +\rmi\epsilon q^+)}U^\dagger(-\pt\!-\qt) \,,
\label{eq:propmp}
\end{equation}
in momentum space.  We note that the color rotation takes place in the
opposite orientation (i.e.\ $U^{-1}=U^\dagger$) to the previous case
because the gluon propagation penetrates into the source from the
opposite side.


\subsection{Propagator and amplitude}

  We now get ready to proceed to evaluating the expanded
propagator~(\ref{eq:prop}).  The most general form is, however, not
necessary for the two-gluon production amplitude.  In coordinate space
only the late time behavior at $x^+\gg0$ and $y^+\gg0$ is relevant to
the production process.  We can actually confirm that the production
amplitude does not have any contribution from the region where $x^+$
and $y^+$ are negative.  In the following argument, thus, we shall
restrict the time arguments as positive $x^+$ and $y^+$ only.


\begin{figure}
\begin{center}
\includegraphics[width=11cm]{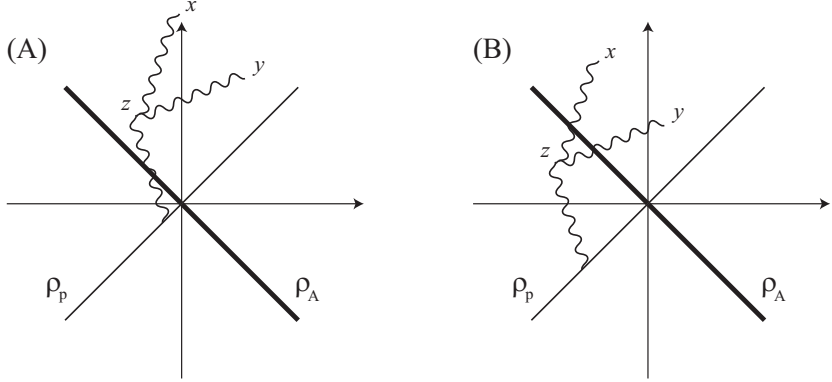}
\end{center}
\caption{Typical diagrams representing the first-order propagator
  which contributes to the two-gluon production amplitude:  (A) One
  gluon splits into two after interaction with the target. (B) Two
  gluons receive interaction with the target.}
\label{fig:G}
\end{figure}


  The production process can be separated into two distinct types
according to the location of the three-point vertex.
Figure~\ref{fig:G} is the diagrammatic representation for them.  We
can distinguish (A) and (B) out of Eq.~(\ref{eq:prop}) by the sign of
$z^+$ so that we can decompose it into
\begin{equation}
 \begin{split}
  \G^{\mu\nu}(x,y) - \G_{(0)}^{\mu\nu}(x,y) =
  & -\int\rmd^4z\, \Delta^{\mu\lambda}(x,z)
   \,\delta\Gamma_{(>)\lambda\sigma}(z)\,\Delta^{\sigma\nu}(z,y) \\
  & -\int\rmd^4z\, \Gpm^{\mu\lambda}(x,z)
   \,\delta\Gamma_{(<)\lambda\sigma}(z)\,\Gmp^{\sigma\nu}(z,y) \,,
\end{split}
\label{eq:prop3}
\end{equation}
where we can forget about $\G_{(0)}^{\mu\nu}(x,y)$ for our purpose.
In the above we have defined,
\begin{equation}
 \delta\Gamma_{(>)\lambda\sigma}(z)\equiv
  \delta\Gamma_{\lambda\sigma}(z)\theta(z^+)\,, \quad
 \delta\Gamma_{(<)\lambda\sigma}(z)\equiv
  \delta\Gamma_{\lambda\sigma}(z)\theta(-z^+)\,.
\end{equation}
This relatively simple form of Eq.~(\ref{eq:prop3}) is, as we stated
above, valid under the condition, $x^+>0$ and $y^+>0$.  The first and
second lines correspond to (A) and (B) of Fig.~\ref{fig:G},
respectively.


\subsection{Evaluating $\Delta^{\mu\lambda}\,
 \delta\Gamma_{(>)\lambda\sigma}\,\Delta^{\sigma\nu}$}

  First we shall explicitly evaluate the contribution at $z^+>0$ shown
in Fig.~\ref{fig:G}~(A).  Since the expression in momentum space is
necessary to acquire the amplitude, let us take the Fourier transform,
\begin{equation}
 \begin{split}
 & \int\rmd^4x\,\rmd^4y\,\rme^{\rmi p\cdot x+\rmi q\cdot y}\,\theta(x^+)
  \theta(y^+) \int\rmd^4z\,\Delta^{\mu\lambda}(x,z)\,
  \delta\Gamma_{(>)\lambda\sigma}(z)\,\Delta^{\sigma\nu}(z,y) \\
 &\qquad\qquad\qquad\qquad
  = -\theta(p^+)\,\theta(q^+)\frac{D^{\mu\lambda}(p)\,
  \delta\Gamma_{(>)\lambda\sigma}(p,q)\,D^{\sigma\nu}(q)}
  {(p^2+\rmi\epsilon)(q^2+\rmi\epsilon)} \,,
 \end{split}
\end{equation}
where  we have used $p^+>0$ and $q^+>0$ in the denominator.  From
Eq.~(\ref{eq:LSZ}) the matrix element
$\langle\p,\lambda,a;\q,\sigma,b|\Omega_\text{in}\rangle$ turns out to
have a contribution as
\begin{equation}
 \langle\p,\lambda,a;\q,\sigma,b|\Omega_\text{in}\rangle_{(>)} =
 - \epsilon^{(\lambda)}_\mu(\p)\,\epsilon^{(\sigma)}_\nu(\q)\,
  \delta\Gamma_{(>)}^{\mu\nu}(p,q) \,.
\end{equation}
Using Eqs.~(\ref{eq:LSZ}), (\ref{eq:sol_ai}), (\ref{eq:sol_a-}), and
(\ref{eq:vertex}) we can figure its explicit form out as
\begin{equation}
 \begin{split}
 & \langle\p,\lambda,a;\q,\sigma,b|\Omega_\text{in}\rangle_{(>)} \\
 &\quad = g\, \epsilon^{(\lambda)}_\mu(\p)\,
  \epsilon^{(\sigma)}_\nu(\q) \int\dk{1}\, T_{(>)}^{\mu\nu}(p,q;\ktone,\kttwo)
  \,\biggl[U(\kttwo)\frac{\rhop(\ktone)}{\ktone^2}
  \biggr]^{ab} \,,
\end{split}
\label{eq:amp_m}
\end{equation}
where $\ktone\!+\kttwo\!=\pt\!+\qt$ and we have used the following
function;
\begin{equation}
 T_{(>)}^{\mu\nu}(p,q;\ktone,\kttwo) \equiv \frac{D^{\mu\lambda'}(p)
  \Gamma_{\lambda'\delta\sigma'}(p,q)D^{\sigma'\nu}(q)}
  {(p+q)^2+\rmi(p^+\!+q^+)\epsilon}\,C^\delta(p+q;\ktone,\kttwo) \,.
\end{equation}
Here, we have defined
\begin{equation}
 \begin{split}
  C^+(p,\ktone,\kttwo) &= 0 \,,\\
  C^-(p,\ktone,\kttwo)
   &= \frac{-2\ktone\cdot\kttwo}{p^++\rmi\epsilon} \,,\\
  C^i(p,\ktone,\kttwo) &= \frac{p^i\ktone^2}
  {(p^++\rmi\epsilon)(p^-+\rmi\epsilon)}-2k_1^i \,,
 \end{split}
\end{equation}
so that we can express the gauge field after the collision in a
concise form as
\begin{equation}
 -p^2\Ac^\mu_{(1>)}(p) = -\rmi\int\dk{1}\, C^\mu(p;\ktone,\kttwo)
  U(\kttwo)\frac{\rhop(\ktone)}{\ktone^2} \,.
\end{equation}


\subsection{Evaluating $\Gpm^{\mu\lambda}\,
 \delta\Gamma_{(<)\lambda\sigma}\,\Gmp^{\sigma\nu}$}

  In the same way we can continue the calculation by taking the
Fourier transformation to obtain,
\begin{equation}
 \begin{split}
 & \int\rmd^4x\,\rmd^4y\,\rme^{\rmi p\cdot x + \rmi q\cdot y}\,
  \theta(x^+)\,\theta(y^+) \int\rmd^4z\,\Gpm^{\mu\lambda}(x,z)\,
  \delta\Gamma_{(<)\lambda\sigma}(z)\,\Gmp^{\sigma\nu}(z,y) \\
 &\qquad\qquad =\int\frac{\rmd^4 k_1}{(2\pi)^4}
  \frac{\rmd^4 k_2}{(2\pi)^4}\,\Gpm^{\mu\lambda}(p,-k_1)\,
  \delta\Gamma_{(<)\lambda\sigma}(k_1,k_2)\,
  \Gmp^{\lambda\nu}(-k_2,q) \,.
 \end{split}
\end{equation}
By substituting the explicit forms given by Eqs.~(\ref{eq:proppm}) and
(\ref{eq:propmp}) we can write the above down explicitly as
\begin{equation}
 \begin{split}
 & \int\frac{\rmd^4 k_1}{(2\pi)^4}\frac{\rmd^4 k_2}{(2\pi)^4}\,
  \Gpm^{\mu\lambda}(p,-k_1)\,\delta\Gamma_{(<)\lambda\sigma}(k_1,k_2)\,
  \Gmp^{\lambda\nu}(-k_2,q) \\
 =\;& \frac{\theta(p^+)}{p^2\!+\rmi\epsilon p^+} \cdot
    \frac{\theta(q^+)}{q^2\!+\rmi\epsilon q^+}
  \int\frac{\rmd^4 k_1}{(2\pi)^4}\frac{\rmd^4 k_2}{(2\pi)^4}\,
  (2\pi)\delta(p^+-k_1^+)\,(2\pi)\delta(-k_2^++q^+) \\
 &\quad \times\frac{2k_1^+ D^{\mu\alpha}(p) D_\alpha^{\;\lambda}(k_1)}
  {(k_1^2+\rmi\epsilon)}\cdot\frac{2k_2^+ D^{\sigma\beta}(k_2)
   D_\beta^{\;\nu}(q)}{(k_2^2+\rmi\epsilon )}  \\
 &\qquad \times U(\pt\!-\ktone)\,
  \delta\Gamma_{(<)\lambda\sigma}(k_1,k_2)\,U^\dagger(\qt\!-\kttwo) \,.
\end{split}
\end{equation}
The poles in $\delta\Gamma_{(<)\lambda\sigma} (k_1,k_2)$ in the complex
$k_1^-$ and $k_2^-$ plane are found on the upper half plane.  We can
then perform the integration with respect to  $k_1^-$ and $k_2^-$
avoiding the pole in $\delta\Gamma_{(<)\lambda\sigma}(k_1,k_2)$, so
that we pick up the contributions
$-(2\pi\rmi)\,(2k_1^+)^{-1}\delta(k_1^-\!-\ktone^2/(2k_1^+))$ from
$1/k_1^2$ and
$-(2\pi\rmi)\,(2k_2^+)^{-1}\delta(k_2^-\!-\kttwo^2/(2k_2^+))$ from
$1/k_2^2$, which leads to
\begin{equation}
 \begin{split}
 & \int \frac{\rmd^4 k_1}{(2\pi)^4}\frac{\rmd^4 k_2}{(2\pi)^4}\,
  \Gpm^{\mu\lambda}(p,-k_1)\,\delta\Gamma_{(<)\lambda\sigma}(k_1,k_2)\,
  \Gmp^{\lambda\nu}(-k_2,q) \\
 =\;& -\frac{\theta(p^+)}{p^2\!+\rmi\epsilon p^+} \cdot
    \frac{\theta(q^+)}{q^2\!+\rmi\epsilon q^+}
    \int\dk{1}\dk{2}\, D^{\mu\alpha}(p) D_\alpha^{\;\lambda}(k_1)
    D^{\sigma\beta}(k_2)D_\beta^{\;\nu}(q) \\
 &\quad \times U(\pt\!-\ktone)\,
  \delta\Gamma_{(<)\lambda\sigma}(k_1,k_2)\,U^\dagger(\kttwo\!-\qt) \,,
\end{split}
\end{equation}
after the integration with respect to $k_1^\pm$ and $k_2^\pm$.  Here
the delta functions impose that $k_1^2=k_2^2=0$, $k_1^+=p^+$,
and $k_2^+=q^+$.  From Eq.~(\ref{eq:LSZ}) the matrix element from this
contribution becomes
\begin{align}
  \langle\p,\lambda,a;\q,\sigma,b|\Omega_\text{in}\rangle_{(<)}
  &=-\epsilon^{(\lambda)}_\mu(\p)\,\epsilon^{(\sigma)}_\nu(\q)
   \int\dk{1}\dk{2}\,D^{\mu\lambda'}(k_1)D^{\sigma'\nu}(k_2) \notag\\
  &\quad\times U(\pt\!-\ktone)\,
   \delta\Gamma_{(<)\lambda'\sigma'}(k_1,k_2)\,U^\dagger(\kttwo\!-\qt) \,,
\end{align}
which simplifies further to be
\begin{equation}
 \begin{split}
  &\langle\p,\lambda,a;\q,\sigma,b|\Omega_\text{in}\rangle_{(<)}
   = g\,\epsilon^{(\lambda)}_\mu(\p)\,\epsilon^{(\sigma)}_\nu(\q)\,
   \int\dk{1}\dk{2} \\
  &\quad\times \biggl[U(\pt\!-\ktone)\,\frac{\rhop(\ktone+\kttwo)}
  {(\ktone+\kttwo)^2}\, U^\dagger(\kttwo\!-\qt)\biggr]^{ab} \!
  T_{(<)}^{\mu\nu}(p,q;\ktone,\kttwo) \,,
 \end{split}
\end{equation}
where we have defined
\begin{equation}
 T_{(<)}^{\mu\nu}(p,q;\ktone,\kttwo)\equiv \frac{D^{\mu\lambda'}(k_1)
  \Gamma_{\lambda'\,i\,\sigma'}(k_1,k_2)D^{\sigma'\nu}(k_2)\,(k_1+k_2)^i}
  {(p^+\!+q^+\!+\rmi\epsilon)(k_1^-\!+k_2^-\!-\rmi\epsilon)} \,.
\end{equation}
Here we recall that $k_1^-=\ktone^2/2p^+$ and $k_2^-=\kttwo^2/2q^+$
from the on-shell condition.

  For later convenience we shall shift the momenta as
\begin{equation}
 \ktone \to \ktone' = \ktone+\kttwo \,,\quad
 \kttwo \to \kt' = \pt-\ktone'+\kttwo \,,
\end{equation}
and omit the prime on the momenta afterward.  Then, the above
amplitude is rewritten into
\begin{align}
 \begin{split}
 &\langle\p,\lambda,a;\q,\sigma,b|\Omega_\text{in}\rangle_{(<)}
  = g\,\epsilon^{(\lambda)}_\mu(\p)\,\epsilon^{(\sigma)}_\nu(\q)
  \int\dk{1}\dk{} \\
 & \times T^{\mu\nu}_{(<)}(p,q;\pt\!-\kt,\,\ktone\!+\kt\!-\pt)\,
  \biggl[U(\kt)\,\frac{\rhop(\ktone)}{\ktone^2}\,
  U^\dagger(\kt\!-\kttwo)\biggr]^{ab} \,.
 \end{split}
\label{eq:amp_p}
\end{align}


\subsection{Total amplitude}


\begin{figure}
\begin{center}
\includegraphics[width=11cm]{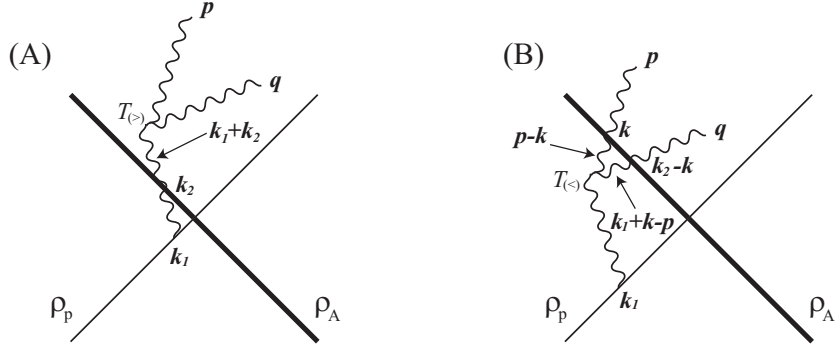}
\end{center}
\caption{Momentum assignment for (A)
$\langle\p,\q|\Omega_\text{in}\rangle_{(>)}$ and (B)
$\langle\p,\q|\Omega_\text{in}\rangle_{(<)}$ respectively.}
\label{fig:mom}
\end{figure}


  After all, the sum of Eqs.~(\ref{eq:amp_m}) and (\ref{eq:amp_p})
amounts to the total amplitude,
\begin{equation}
 \begin{split}
  &\langle\p,\lambda,a;\q,\sigma,b|\Omega_\text{in}\rangle
   = g\,\epsilon^{(\lambda)}_\mu(\p)\,\epsilon^{(\sigma)}_\nu(\q)
   \int\dk{1}\,\frac{\rhop^c(\ktone)}{\ktone^2} \\
  & \times\biggl[ T_{(>)}^{\mu\nu}(p,q;\ktone,\kttwo)\,\ta^d\,U^{dc}(\kttwo)
   +\int\dk{}\\
  &\qquad\times T_{(<)}^{\mu\nu}(p,q;\pt\!-\kt,\,\ktone\!+\kt\!-\pt)
   \,U(\kt) \ta^c U^\dagger(\kt\!-\kttwo)\biggr]^{ab} \,.
 \end{split}
\label{eq:total_amp}
\end{equation}
Figure~\ref{fig:mom} shows the momentum assignment overlaid on
Fig.~\ref{fig:G}.  In the above expression the SU($\Nc$) algebra in
the adjoint representation is denoted by $\ta$.  Let us now confirm
that this results are certainly reduced to zero in the limit of
vanishing target source, that is,
$U(\kt)\to (2\pi)^2\delta^{(2)}(\kt)$, for consistency check.

  Then, the quantity inside the square brackets is
\begin{equation}
 \ta^c \bigl[T_{(>)}^{\mu\nu}(p,q;\pt\!+\qt,\zerot)
  + T_{(<)}^{\mu\nu}(p,q;\pt,\qt)\bigr] \,.
\label{eq:zero}
\end{equation}
After some calculations we can find that
\begin{equation}
 \begin{split}
  T_{(>)}^{\mu\nu}(p,q;\pt\!+\qt,\zerot) &= -T_{(<)}^{\mu\nu}(p,q;\pt,\qt)\\
  &= \frac{D^{\mu\lambda}(p)\Gamma_{\lambda i\sigma}(p,q)D^{\sigma\nu}(q)\,
   (p^i\!+q^i)}{(p^+\!+q^+)(p^-\!+q^-)} \,,
 \end{split}
\end{equation}
which makes Eq.~(\ref{eq:zero}) canceled as anticipated.

  For later convenience let us write Eq.~(\ref{eq:total_amp}) in a
slightly different way.  Noting that,
\begin{equation}
 \int\dk{}\,U(\kt)\ta^c U^\dagger(\kt\!-\kttwo)
  = \ta^d U^{dc}(\kttwo) \,,
\end{equation}
we can rewrite the first and second terms altogether as
\begin{equation}
 \begin{split}
  &\langle\p,\lambda,a;\q,\sigma,b|\Omega_\text{in}\rangle \\
  & = g\,\epsilon^{(\lambda)}_\mu(\p)\,\epsilon^{(\sigma)}_\nu(\q)
   \int\dk{1}\,\frac{\rhop^c(\ktone)}{\ktone^2}\int\dk{}
   \biggl[U(\kt) \ta^c U^\dagger(\kt\!-\kttwo)\biggr]^{ab} \\
  & \qquad\times\biggl[ T_{(>)}^{\mu\nu}(p,q;\ktone,\kttwo)
   + T_{(<)}^{\mu\nu}(p,q;\pt\!-\kt,\,\ktone\!+\kt\!-\pt) \biggr] \,.
   \end{split}
\label{eq:total_amp2}
\end{equation}
This is a useful form because, as we will face soon below, the sum of
two terms in the angle brackets has partial cancellation in some
cases.


\section{Rapidity Correlation}

  Now that we have worked out the amplitude, its squared quantity
immediately leads us to an expression for the correlation function
shown in Fig.~\ref{fig:production}~(C) (i.e.\ connected diagram), that
is,
\begin{align}
 &\biggl\langle \frac{\rmd\Ng}{\rmd y_1\rmd y_2}
  \biggr\rangle_{\text{conn}} \!\!\!\!
  = \frac{g^2}{16\pi^2}\int\dpt{}\dqt{}\dk{1}\dprimek{1}
  \frac{\langle\rhop^a(\ktone)\rhop^{a'\ast}(\ktone')\rangle}
  {\ktone^2{\ktone'}^2} \notag\\
 &\times \biggl\{\tr[T_{(>)}\cdot T_{(>)}^\dagger]
  \bigl\langle U^{ba}(\kttwo) U^{\dagger a'b'}(\kttwo')
  \tr[\ta^b \ta^{b'}]\bigr\rangle \notag\\
 &\qquad +\int\dk{}\, \tr[T_{(<)}\cdot T_{(>)}^\dagger]
  \bigl\langle\tr\bigl[U(\kt)\ta^a U^\dagger(\kt\!-\kttwo)
  \ta^{b'}\bigr] U^{\dagger a'b'}(\kttwo')\bigr\rangle \notag\\
 &\qquad +\int\dprimek{}\, \tr[T_{(>)}\cdot T_{(<)}^\dagger]
  \bigl\langle U^{ba}(\kttwo)\,\tr\bigl[\ta^b
  U(\kt'\!-\kttwo')\ta^{a'} U^\dagger(\kt')\bigr]\bigr\rangle \notag\\
 &\qquad +\int\dk{}\dprimek{}\,\tr[T_{(<)}\cdot T_{(<)}^\dagger]
  \notag\\
 &\qquad\quad \times \bigl\langle\tr\bigl[ U(\kt)\ta^a
   U^\dagger(\kt\!-\kttwo)\, U(\kt'\!-\kttwo')\ta^{a'}
   U^\dagger(\kt')\bigr]\bigr\rangle \biggr\} \,,
\label{eq:final}
\end{align}
where we have suppressed the momentum arguments for
$T_{(>)}^{\mu\nu}\equiv T_{(>)}^{\mu\nu}(p,q;\ktone,\kttwo)$,
$T_{(>)}^{\dagger\mu\nu}\equiv T_{(>)}^{\dagger\mu\nu}(p,q;\ktone',\kttwo')$,
$T_{(<)}^{\mu\nu}\equiv T_{(<)}^{\mu\nu}(p,q;\pt\!-\kt,\,\ktone\!+\kt\!-\pt)$,
and
$T_{(<)}^{\dagger\mu\nu}\equiv  T_{(<)}^{\dagger\mu\nu}(p,q;\pt\!-\kt',\,
 \ktone'\!+\kt'\!-\pt)$.  We note that $\kttwo$ and $\kttwo'$ are
defined through
$\ktone\!+\kttwo\!=\ktone'\!+\kttwo'\!=\pt\!+\qt$.  These are our
final results.  We present the results in a somewhat more intuitive
form in terms of the gluon distribution functions in
Appendix~\ref{sec:distribution}.  The evaluation for the Wilson line
average is supplemented in Appendix~\ref{sec:Wilson}.

  In the rest of this section we further explore the rapidity
correlation out of Eq.~(\ref{eq:final}) in some special situations.
First, let us consider the remaining correlation which is rapidity
independent in the long range.  We assumed $|\dely|<1/\alpha_s$ to
avoid the quantum evolution effect, as we mentioned in Introduction.
Now we shall consider the correlation in the rapidity region
satisfying $1\ll|\dely|<1/\alpha_s$, so that the eikonal approximation
can work for emitted gluons.  Then, we set $y_2\gg 1$ and $y_1\ll -1$
and regard $q^+$ and $p^-$ as much greater than other scales.  Then,
$q^+$ is so large that the gluon with $q^+$ cannot propagate
longitudinally due to suppression by $1/q^+$ in Eq.~(\ref{eq:D}).  As
a result, the longitudinal component of the gauge field
$\mathcal{A}^-$ is to be neglected in the eikonal limit.

  In this approximation it is immediate to see;
\begin{equation}
 \begin{split}
  & (p+q)^2 \approx 2q^+p^- \,,\\
  & C^-(p+q,\ktone,\kttwo) \approx 0 \,,\quad
  C^i(p+q,\ktone,\kttwo) \approx -2k^i \,,\\
  & \Gamma_{\mu \lambda\nu}(p,q) \approx q^+
   (-g_{+\lambda}g_{\mu\nu} + 2g_{+\mu}g_{\nu\lambda}) \,,
 \end{split}
\end{equation}
where we have dropped two terms from $\Gamma_{\mu\lambda\nu}(p,q)$
because no finite contribution arises whenever either of $\mu$,
$\lambda$, or $\nu$ takes $+$.  After some calculations with using the
on-shell conditions we can find $T_{(>)}^{ij}$ and  $T_{(<)}^{ij}$ to
be rewritten as
\begin{align}
 & T_{(>)}^{ij}(p,q;\ktone,\kttwo) \approx \frac{4p^ik_1^j}{\pt^2}
  = -\frac{4\pi\ktone^2}{g^2}f^i(\pt)f^j(\ktone) \,,
\label{eq:long_>}\\
 & T_{(<)}^{ij}(p,q;\pt\!\!-\kt,\ktone\!\!+\kt\!\!-\pt) \approx
  \frac{-4(p\!-\!k)^ik_1^j}{(\pt\!-\kt)^2}
  = \frac{4\pi\ktone^2}{g^2}f^i(\pt\!\!-\kt)\,f^j(\ktone) \,,
\label{eq:long_<}
\end{align}
where we define $f^i$ following the notation in
Ref.~\cite{Baier:2005dv} by
\begin{equation}
 f^i(\pt)=\frac{-\rmi g}{\sqrt{\pi}} \frac{p^i}{\pt^2} \,.
\end{equation}
We note that $f^i(\ktone)\rhop(\ktone)$ is nothing but the
perturbative Weizs\"acker-Williams gluon field.  From a different
point of view, the above together with $\epsilon_i^{(\lambda)}(\p)$
can be interpreted as coming from the amplitude of the soft gluon
emission in the color dipole picture.

  After all, the total amplitude becomes as simple as
\begin{equation}
 \begin{split}
  \langle\p,i,a;\q,j,b|\Omega_\text{in}\rangle
   &= -\frac{4\pi}{g} \int\dk{1}\,f^j(\ktone)\,\rhop^c(\ktone)
   \biggl[ f^i(\pt)\ta^d\,U^{dc}(\kttwo) \\
  &\qquad -\int\dk{}\, f^i(\pt\!-\kt) U(\kt) \ta^c
   U^\dagger(\kt\!-\kttwo)\biggr] \,.
 \end{split}
\label{eq:eikonal_amp}
\end{equation}
This production amplitude given in coordinate space exactly coincides
with the third term of Eq.~(5.8) in Ref.~\cite{Baier:2005dv} except
for the normalization.  We should remark that
Eq.~(\ref{eq:eikonal_amp}) is independent of the rapidity difference
$\dely$ as we neglect $\rme^{-\dely}$ under the condition that only
$q^+$ and $p^-$ are dominating.

  Next, we will consider a different situation with the rapidity
dependence handled easily.  Let $\pt$ and $\qt$ be much bigger than
$\ktone$, $\kttwo$, and $\kt$.  This means that $\pt$ and $\qt$ belong
to the hard scale as compared to the saturation scale which
characterizes the gluon distribution inside the target nucleus.  We
note that this situation is the case if we are interested in the
di-jet correlations.  We shall further limit our calculation here to
the back-to-back kinematics, i.e.\
\begin{equation}
 \pt + \qt = \ktone\! + \kttwo = \zerot \,,
\end{equation}
for simplicity.

  As a matter of fact, the back-to-back kinematics significantly
reduces complexity enabling us to perform explicit calculations with a
reasonable effort.  Without any expansion we can express $T_{(>)}$ and
$T_{(<)}$ as
\begin{align}
 & T_{(>)}^{ij}(p,q;\ktone,\kttwo) \notag\\
 & = \frac{\bigl[\ktone^2\tanh(\dely/2)+2\ktone\!\!\cdot\pt\bigl]
  g^{ij} + 2k_1^j p^i(1+\rme^{-\dely}) + 2k_1^i p^j(1+\rme^{\dely})}
  {\pt^2 (1+\cosh\dely)} \,,
\label{eq:t>}\\
 & T_{(<)}^{ij}(p,q;\pt\!-\kt,\ktone\!+\kt\!-\pt) \notag\\
 & = -\frac{\ktone\!\!\cdot\!(2\pt\!\!\!-2\kt\!\!\!-\ktone) g^{ij}
  \!\!+\! 2k_1^j (p^i\!\!-\!k^i)(1\!\!+\!\rme^{-\dely})
  \!+\! 2k_1^i (p^j\!\!-\!k^j\!-\!k_1^j)(1\!\!+\!\rme^{\dely})}
  {(\pt\!-\kt)^2 (1+\rme^{-\dely})/2 + (\pt\!-\kt\!-\ktone)^2
  (1+\rme^{\dely})/2} .
\label{eq:t<}
\end{align}
Here it is easy to make sure that the above two expressions are to be
transmuted to Eqs.~(\ref{eq:long_>}) and (\ref{eq:long_<}) in the
limit of $\dely\to\infty$.  It is also manifest that $T_{(>)}+T_{(<)}$
is vanishing up to the first order in terms of $\ktone$ and $\kt$.
Let us henceforth look more into the rapidity dependence to grasp a
feeling from the above results.

  We expand Eqs.~(\ref{eq:t>}) and (\ref{eq:t<}) in terms of $\ktone$
and $\kt$ keeping the quadratic order.  After tedious but
straightforward calculations we can reach;
\begin{equation}
 T^{ij} = T_{(>)}^{ij}+T_{(<)}^{ij}
  \approx \frac{2\ktone\!\cdot\kbar g^{ij}
  + 2k_1^j \overline{k}^i (1+\rme^{-\dely})
  + 2k_1^i \overline{k}^j (1+\rme^{\dely})}
  {\pt^2(1+\cosh\dely)} \,,
\end{equation}
where we define
\begin{equation}
 \kbar = \ktilde - 2(\phat\!\cdot \ktilde)\phat \,,\qquad
 \ktilde = \kt + \frac{1}{1+\rme^{-\dely}}\ktone \,,
\end{equation}
and $\phat=\pt/|\pt|$.  The system having translational symmetry in
the transverse plane  constrains as $\ktone'=\ktone$, but not for
$\kt$ and $\kt'$.  The squared quantity of the above matrix takes a
relatively simple form of
\begin{equation}
 \tr[T\cdot T^\dagger] = \frac{8\bigl[
  (\ktone\!\!\cdot\kbar)(\ktone\!\!\cdot\kbar')
  + 2\ktone^2 (\kbar\!\!\cdot\kbar')
  \cosh\dely\,(1+\cosh\dely)\bigr]}{(\pt^2)^2 (1+\cosh\dely)^2} \,.
\end{equation}
Here the angular average with respect to $\phat$ can take us to make
further simplification, so that we finally get
\begin{equation}
 \tr[T\cdot T^\dagger] = \frac{4\ktone^2 (\ktilde\!\cdot\ktilde')}
  {(\pt^2)^2} \cdot \biggl(\frac{1+2\cosh\dely}{1+\cosh\dely}\biggr)^2 \,.
\end{equation}
Here we should not forget that $\ktilde$ and $\ktilde'$ have the
rapidity dependence too.  By shifting $\kt\to\kt-\ktone/2$ and
$\kt'\to\kt'-\ktone/2$, the final expression for the back-to-back
two-gluon production from the connected diagram is
\begin{align}
 & \biggl\langle \frac{\rmd\Ng}{\rmd|\pt|\rmd|\qt|\rmd\theta
  \rmd y_1\rmd y_2} \biggr|_{\qt=-\pt}\biggr\rangle_{\text{conn}}
  \!\!\!
  = \frac{1}{\pi R_\text{p}^2} \frac{4\alpha_{\text{s}}}{(\Nc^2\!-\!1)
  (2\pi)^5\pt^2}\int\dk{1} \varphi_{\text{p}}(\ktone) \notag\\
 & \times\int\dk{}\dprimek{} \Bigl[\kt\!\cdot\kt'
  + \frac{1}{4}\bigl(\tanh(\dely/2)\bigr)^2 \ktone^2 \Bigr]\cdot
  \biggl(\frac{1+2\cosh\dely}{1+\cosh\dely}\biggr)^2
\label{eq:y_dep}\\
 & \times\bigl\langle\tr[U(\kt\!\!-\!\ktone\!/2)\ta^a
  U^\dagger(\kt\!\!+\!\ktone\!/2)\,
  U(\kt'\!\!+\!\ktone\!/2)\ta^a U^\dagger(\kt'\!\!-\!\ktone\!/2)]
  \bigr\rangle \,, \notag
\end{align}
where $\theta$ is the angle between $\qt$ and $\pt$, and we have used
the gluon distribution function defined in Eq.~(\ref{eq:phip}) in
Appendix~\ref{sec:distribution}.  The transverse area
$\pi R_{\text{p}}^2$ originates from our definition of the gluon
distribution function.

The expectation value of the Wilson line correlator includes contributions
from all twists, which is numerically evaluated.
If we make a leading twist approximation, we can analytically evaluate
the Wilson line correlation, which is simply obtained as
\begin{align}
 & \bigl\langle\tr[U(\kt\!\!-\!\ktone\!/2)\ta^a
  U^\dagger(\kt\!\!+\!\ktone\!/2)\,
  U(\kt'\!\!+\!\ktone\!/2)\ta^a U^\dagger(\kt'\!\!-\!\ktone\!/2)]
  \bigr\rangle \, \notag
  \\
  &\approx 
  \frac{4\alpha_s\varphi_\text{A}(\ktone)}{\ktone^2}\Nc^2\bigl(
 \delta^{(2)}(\kt\!\!+\!\frac{\ktone}{2})\delta^{(2)}
  (\kt'\!\!+\!\frac{\ktone}{2})
  +\delta^{(2)}(\kt\!\!-\!\frac{\ktone}{2})
  \delta^{(2)}(\kt'\!\!-\!\frac{\ktone}{2}) \notag\\
  &\;\;\;\;  -\frac{1}{2}\delta^{(2)}(\kt\!\!+\!\frac{\ktone}{2})
  \delta^{(2)}(\kt'\!\!-\!\frac{\ktone}{2})
  -\frac{1}{2}\delta^{(2)}(\kt'\!\!+\!\frac{\ktone}{2})
  \delta^{(2)}(\kt\!\!-\!\frac{\ktone}{2}) \bigr) \;,
\end{align}
where we have defined $\varphi_\text{A}$ to mean the gluon
distribution function of the nucleus.  In this approximation, we
obtain the correlation function as
\begin{align}
 &\biggl\langle \frac{\rmd\Ng}{\rmd|\pt|\rmd|\qt|\rmd \theta
  \rmd y_1\rmd y_2} \biggr|_{\qt=-\pt}\biggr\rangle_{\text{conn}} \!\!\!
 =\frac{1}{\pi R_\text{p}^2} \frac{4\alpha_s^2\Nc^2}
  {(\Nc^2\!-\!1)(2\pi)^5\pt^2}\int\dk{1}\notag \\
 &\qquad\times \varphi_{\text{p}}(\ktone)\,\varphi_{\text{A}}(\ktone) 
 \Bigl\{3+\bigl(\tanh(\dely/2)\bigr)^2 \Bigr\}
  \biggl(\frac{1+2\cosh\dely}{1+\cosh\dely}\biggr)^2 \;.
  \label{eq:rapiditydependence}
\end{align}

\begin{figure}
 \begin{center}
 \includegraphics[width=9cm]{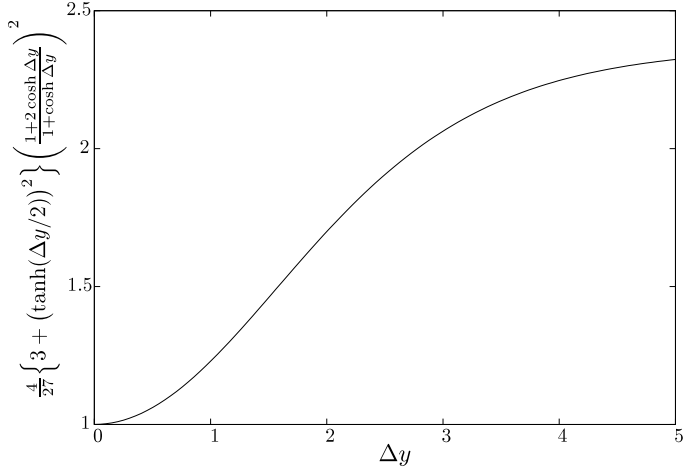}
 \end{center}
 \caption{Dependence on $\dely$ in the leading-twist approximation.}
 \label{fig:function}
\end{figure}

Although the expression in Eq.~(\ref{eq:rapiditydependence}) has a
$k_\perp$-factorized form, we cannot write the two-gluon production in
general into such a $k_\perp$-factorized expression even in the pA
case~\cite{Blaizot:2004wu,Fujii:2006ab}.  In this back-to-back
kinematics, the rapidity dependence is simply factorized as
\begin{align}
 &\biggl\langle \frac{\rmd\Ng}{\rmd|\pt|\rmd|\qt|\rmd
  \theta\rmd y_1\rmd y_2} \biggr|_{\qt=-\pt}\biggr\rangle_{\text{conn}}\!\!\!
  = \biggl\langle \frac{\rmd\Ngg}{\rmd|\pt|\rmd|\qt|\rmd
  \theta\rmd y_1\rmd y_2} \biggr|_{\qt=-\pt,\dely=0}\biggr\rangle_{\text{conn}}
  \notag\\
 &\qquad\qquad\qquad \times\frac{4}{27}
  \Bigl\{3+\bigl(\tanh(\dely/2)\bigr)^2 \Bigr\}
  \biggl(\frac{1+2\cosh\dely}{1+\cosh\dely}\biggr)^2 \;.
\end{align} 
The correlation function is a monotonously increasing function of
$\dely$.  In Fig.~\ref{fig:function}, we show the rapidity correlation
associated with the two-gluon production in the back-to-back case.  It
is remarkable that the correlation is enhanced as the rapidity
difference increases, which is valid as long as
$\dely<1/\alpha_\text{s}$.   At asymptotically large rapidity
difference with $\dely\to\infty$, if the quantum evolution between
$\dely$ is neglected, the correlation function exactly reproduces
the result in the eikonal limit, which is in agreement with
Ref.~\cite{Leonidov:1999nc}.

\section{Conclusions}

  We explicitly calculated the two-gluon correlation in the pA
collision using the framework of the Color Glass Condensate and the
LSZ reduction formula.  We specifically chose a connected diagram,
which is the leading order of the coupling $g$ and the weak color
source, and could be the only contribution to produce the intrinsic
dependence on the longitudinal or rapidity separation between emitted
gluons.  To accomplish the calculation we wrote down the classical
gluon field associated with the dilute and dense color sources and put
it to the background propagator.

  Our final results encompass the rapidity dependence that is not
taken into account in the eikonal limit, but they are too complicated
to perceive the qualitative behavior as a function of the rapidity
separation.  We have to wait for numerical evaluation, which we plan
to do as a natural extension of this work, in order to extract
phenomenological physics implication from our analytical expressions.

  In this paper we proceeded to analyzing the rapidity dependence in
the special cases.  First, we confirmed that our expressions are
certainly reduced to the known expression given in
Ref.~\cite{Baier:2005dv} in the eikonal limit.  Next, we relaxed this
limit so as to allow for the rapidity separation under the condition
that the momenta carried by emitted gluons are hard as compared to the
gluon distribution in the projectile and target hadrons.  In this case
we found a fairly concise formula for the two-gluon production in the
back-to-back kinematics.  Our results imply that the dependence on the
rapidity separation is only moderate and there arise no exponentially
damping correlations.

  In fact, our finding is the other way around;  the connected diagram
enhances the long-range correlations in rapidity.  That is, the
contribution to the multiplicity from the connected diagram gets
larger with increasing rapidity separation.  Although this might sound
curious at first, this could be possible because we consider two
gluons split from one gluon.  The long-range correlations in this case
simply signifies that more gluons come out with larger longitudinal
angle.  In fact, the three-point vertex in QCD contains
$p^+/q^+=\rme^{\dely}$ which surpasses the energy denominator.  We
remind that our tree-level description breaks down eventually for
$\dely>1/\alpha_{\text{s}}$ due to quantum effects.

  It is quite interesting to see such long-range correlations in
rapidity and this observation seems to be consistent with the ridge
phenomenon observed at
RHIC~\cite{Majumder:2006wi,Dumitru:2008wn,Gavin:2008ev}.  We cannot,
of course, make any convincing statement about the near-side jet
correlations from our rough analyses limited to the back-to-back
kinetics.  It is indispensable to perform the numerical integration to
quantify the two-gluon production.  In the near future we are planning
to include the disconnected diagram, which is much simpler than the
connected one we manipulated here, to discuss not only the rapidity
dependence but also the azimuthal angle dependence and their interplay
in this framework \cite{hidaka:2008}.


  We thank Larry McLerran for encouraging us to initiate this
problem.  We thank Fran\c{c}ois Gelis and Raju Venugopalan for
discussions. We also thank Fran\c{c}ois Fillion-Gourdeau for useful
comments.  K.~F.\ is supported by Japanese MEXT grant No.\ 20740134
and also supported in part by Yukawa International Program for Quark
Hadron Sciences (YIPQS).
This research is supported in part by RIKEN BNL
Research Center and the US\ Department of Energy under cooperative
research agreement \#DE-AC02-98CH10886.
\appendix


\section{Appendix -- Gluon Distributions}
\label{sec:distribution}

  To make the final expression~(\ref{eq:final}) more comprehensible in
an intuitive way let us make use of the notation by the parton
distribution functions along the similar line as
Ref.~\cite{Blaizot:2004wu}.  We first define the proton unintegrated
gluon distribution;
\begin{equation}
 \begin{split}
  &\bigl\langle \rhop^a(\ktone)\rhop^{a'\ast}(\ktone') \bigr\rangle \\
  &= \frac{\delta^{aa'} \,\ktone^2}{\pi(\Nc^2-1)}
    \int\rmd^2\Xt\, \rme^{\rmi(\ktone\!-\ktone')\cdot\Xt}
  \frac{\rmd\varphi_{\text{p}}\bigl(\ktone
  \big|\Xt\!\bigr)}{\rmd^2\Xt} \,,
\end{split}
\label{eq:phip}
\end{equation}
where we have assumed $\ktone\approx\ktone'$.
The above equality could be understood as the definition for
$\varphi_{\text{p}}(\kt|\Xt)$.  It is a straightforward generalization
to proceed to the definition for the two-point correlator, that is,
\begin{equation}
 \begin{split}
 &\delta^{aa'}\bigl\langle U^{ba}(\kttwo)\,U^{\dagger a'b'}(\kttwo')
  \tr[\ta^b \ta^{b'}]\bigr\rangle
 = \Nc\bigl\langle \tr\bigl[U(\kttwo)\,U^\dagger(\kttwo')\bigr]
  \bigr\rangle \\
 &= \frac{g^2\Nc^2}{\pi\kttwo^2} \int\rmd^2\Xt\,
  \rme^{i(\kttwo\!-\kttwo')\cdot\Xt}\,
  \frac{\rmd\varphi^{g,g}_A(\kttwo|\Xt\!-\bb)}{\rmd^2\Xt} \,.
 \end{split}
\end{equation}
We should note that $\bb$ is the impact parameter that specifies the
relative separation from the proton gluon distribution in the
transverse plane.  The three-point correlator goes on to be
\begin{equation}
 \begin{split}
 &\delta^{aa'}\bigl\langle\tr\bigl[U(\kt)\ta^a U^\dagger(\kt\!-\kttwo)
  \ta^{b'}\bigr]U^{\dagger a'b'}(\kttwo')\bigr\rangle \\
 &= \frac{g^2\Nc^2}{\pi\kttwo^2} \int\rmd^2\Xt\,
  \rme^{i(\kttwo\!-\kttwo')\cdot\Xt}
  \frac{\rmd\varphi^{gg,g}_A(\kt,\kttwo\!-\kt;\kttwo|\Xt\!-\bb)}
  {\rmd^2\Xt} \,.
 \end{split}
\end{equation}
The four-point correlator is
\begin{equation}
 \begin{split}
 &\delta^{aa'}\bigl\langle\tr\bigl[U(\kt)\ta^a U^\dagger(\kt\!-\kttwo)
  U(\kt'\!-\kttwo')\ta^{a'} U^\dagger(\kt')\bigr]\bigr\rangle \\
 &= \frac{g^2\Nc^2}{\pi\kttwo^2} \int\rmd^2\Xt\,
  \rme^{i(\kttwo\!-\kttwo')\cdot\Xt}
  \frac{\rmd\varphi^{gg,gg}_A(\kt,\kttwo\!-\kt;\kt',
  \kttwo\!-\kt'|\Xt-\bb)}{\rmd^2\Xt} \,.
 \end{split}
\end{equation}
Here we assume that the distribution functions are approximately
homogeneous in transverse space, which is often the case in the
McLerran-Venugopalan model treatment based on the Gaussian
distribution for the color source.  Then, the integration over the
transverse plane gives rise to the momentum delta functions, which
simplifies the final results to be
\begin{align}
  &\biggl\langle\frac{\rmd\Ng}{\rmd y_1\rmd y_2}
   \biggr\rangle_{\text{conn}} \!\!\!= \frac{1}{\pi R_\text{p}^2}
   \frac{\alpha^2_s\Nc^2}
   {(\Nc^2-1)\pi^2}\int\dpt{}\dqt{}\dk{1} \;
   \frac{\varphi_{\text{p}}(\ktone)}{\ktone^2\kttwo^2} \notag\\
  &\times \biggl\{\tr[T_{(>)}\cdot T_{(>)}^\dagger]\,\varphi^{g,g}_A(\kttwo)
   +\int\dk{}\, \tr[T_{(<)}\cdot T_{(>)}^\dagger]\,
   \varphi^{gg,g}_A(\kt,\kttwo\!-\kt;\kttwo) \notag\\
  &\qquad +\int\dprimek{}\,\tr[T_{(>)}\cdot T_{(<)}^\dagger]\,
   \varphi^{gg,g}_A(\kt,\kttwo\!-\kt;\kttwo) \\
  &\qquad +\int\dk{}\dprimek{}\, \tr[T_{(<)}\cdot T_{(<)}^\dagger]\,
   \varphi^{gg,gg}_A(\kt,\kttwo\!-\kt;\kt',\kttwo\!-\kt') \biggr\} \,,
   \notag
\end{align}
where $R_\text{p}$ is the transverse radius of the proton.  It is
necessary to calculate the gluon distribution functions for numerical
evaluation and we can utilize the McLerran-Venugopalan model.
Appendix~\ref{sec:Wilson} is devoted to the explanation of how to
compute the average of the Wilson lines.


\section{Appendix -- Average of the Wilson Lines}
\label{sec:Wilson}

  In this appendix, we evaluate the Gaussian average for the two-,
three- and four-point functions in terms of the Wilson line in the
adjoint representation.  They are all necessary to compute the gluon
distribution functions in the McLerran-Venugopalan model.  For this
purpose it is enough to calculate the four-point function because the
two- and three-point functions are readily deduced by the appropriate
contraction from the four-point function as
\begin{equation}
 \Nc\bigl\langle\tr[U(\bx_{1\perp})U^\dagger(\bx_{2\perp})]\bigr\rangle
  = \bigl\langle\tr[U(\bx_{1\perp})\ta^a U^\dagger(\bx_{3\perp})
  U(\bx_{3\perp})\ta^a U^\dagger(\bx_{2\perp})]\bigr\rangle\,,
\end{equation}
and
\begin{equation}
 \begin{split}
  & \bigl\langle\tr[U(\bx_{1\perp})\ta^a U^\dagger(\bx_{2\perp})\ta^b]
   U^{\dagger ab}(\bx_{3\perp}) \bigr\rangle \\
  &\qquad =\bigl\langle \tr[U(\bx_{1\perp})\ta^a U^\dagger(\bx_{2\perp})
   U(\bx_{3\perp})\ta^a U^\dagger(\bx_{3\perp})]\bigr\rangle \,.
 \end{split}
\end{equation}

  Calculating the generic four-point function in the adjoint
representation is a tedious task, though it is not impossible.
Instead, we thus adopt the large-$\Nc$ expansion here.  We can express
the adjoint Wilson line using the fundamental Wilson lines as
\begin{equation}
 U(\bx_{i\perp})_{\beta\alpha}=2\tr[\tf^\beta U_{\text{F}}(\bx_{i\perp})
  \tf^\alpha U_{\text{F}}^\dagger(\bx_{i\perp})] \,,
\label{eq:adjointWilsonline}
\end{equation}
where $U_{\text{F}}(\bx_{i\perp})$ is the fundamental Wilson line and
$\tf$ is the generator in the fundamental representation.  Using
Eq.~(\ref{eq:adjointWilsonline}) we can then rewrite the four-point
function in the adjoint representation as follows;
\begin{align}
 & \bigl\langle \tr[U(\bx_{1\perp})\ta^a U^\dagger(\bx_{2\perp})
  U(\bx_{3\perp})\ta^a U^\dagger(\bx_{4\perp})]\bigr\rangle \notag\\
 & =4{\ta^a}_{\alpha_1\alpha_2}{\ta^a}_{\alpha_3\alpha_4}
     {\tf^{\alpha_1}}_{a_1\bar{a}_1}{\tf^{\alpha_2}}_{a_2\bar{a}_2}
     {\tf^{\alpha_3}}_{a_3\bar{a}_3}{\tf^{\alpha_4}}_{a_4\bar{a}_4} \\
 & \quad \times 4\delta_{\beta_1\beta_4}\delta_{\beta_2\beta_3}
  {\tf^{\beta_1}}_{\bar{b}_1b_1}{\tf^{\beta_2}}_{\bar{b}_2b_2}
  {\tf^{\beta_3}}_{\bar{b}_3b_3}{\tf^{\beta_4}}_{\bar{b}_4b_4}
  \bigl\langle \prod_{i=1}^4 U_{\text{F}b_ia_i}(\bx_{i\perp})
  U^\ast_{\text{F}\bar{b}_i\bar{a}_i}(\bx_{i\perp}) \bigr\rangle \,. \notag
\end{align}
In the same way as the present authors did in
Ref.~\cite{Fukushima:2007dy} we shall introduce the ``initial state''
and the ``finial state'' as
\begin{align}
 & \langle a_1,a_2,a_3,a_4,\bar{a}_1,\bar{a}_2,\bar{a}_3,\bar{a}_4|
  \text{initial}\rangle = 4{\ta^a}_{\alpha_1\alpha_2}{\ta^a}_{\alpha_3\alpha_4}
  {\tf^{\alpha_1}}_{a_1\bar{a}_1}{\tf^{\alpha_2}}_{a_2\bar{a}_2}
  {\tf^{\alpha_3}}_{a_3\bar{a}_3}{\tf^{\alpha_4}}_{a_4\bar{a}_4} \notag\\
 & = \frac{1}{2}\bigl(\delta_{a_1\bar{a}_4}\delta_{a_2\bar{a}_1}
  \delta_{a_3\bar{a}_2}\delta_{a_4\bar{a}_3}+\delta_{a_1\bar{a}_2}
  \delta_{a_2\bar{a}_3}\delta_{a_3\bar{a}_4}\delta_{a_4\bar{a}_1} \notag\\
 &\qquad\qquad -\delta_{a_1\bar{a}_3}\delta_{a_2\bar{a}_1}\delta_{a_3\bar{a}_4}
  \delta_{a_4\bar{a}_2}-\delta_{a_1\bar{a}_2}\delta_{a_2\bar{a}_4}
  \delta_{a_3\bar{a}_1}\delta_{a_4\bar{a}_3} \bigr)\,, \\
 & \langle \text{final}|b_1,b_2,b_3,b_4,\bar{b}_1,\bar{b}_2,
  \bar{b}_3,\bar{b}_4\rangle = 4 \delta_{\beta_1\beta_4}
  \delta_{\beta_2\beta_3} {\tf^{\beta_1}}_{\bar{b}_1b_1}{\tf^{\beta_2}}_{\bar{b}_2b_2}
  {\tf^{\beta_3}}_{\bar{b}_3b_3}{\tf^{\beta_4}}_{\bar{b}_4b_4} \notag\\
 & = \delta_{b_1\bar{b}_4}\delta_{b_2\bar{b}_3}\delta_{b_3\bar{b}_2}
  \delta_{b_4\bar{b}_1} +\frac{1}{\Nc^2}\delta_{b_1\bar{b}_1}
  \delta_{b_2\bar{b}_2}\delta_{b_3\bar{b}_3}\delta_{b_4\bar{b}_4} \notag\\
 &\qquad\qquad -\frac{1}{\Nc}\bigl(\delta_{b_1\bar{b}_1}\delta_{b_2\bar{b}_3}
  \delta_{b_3\bar{b}_2}\delta_{b_4\bar{b}_4} + \delta_{b_1\bar{b}_4}
  \delta_{b_2\bar{b}_2}\delta_{b_3\bar{b}_3}\delta_{b_4\bar{b}_1} \bigr) \,,
\end{align}
and also define the ``Hamiltonian'' composed of the free part, $H_0$,
and the interaction $V$;
\begin{align}
 H_0 &= \frac{\Qs'^2}{\Nc} L(0,0)\biggl[\sum_{i=1}^4
  ({\tf}_i^a-{\tf}_{\bar{i}}^{a\ast})\biggr]^2 \,,\\
 V &= -\frac{\Qs'^2}{\Nc}\biggl\{\sum_{i>j}^4
  \bigl[ {\tf}^a_i\,{\tf}^a_j\,\varGamma(\bx_{i\perp},\bx_{j\perp})
  +{\tf}^{a\ast}_{\bar{i}}\,{\tf}^{a\ast}_{\bar{j}}\,
  \varGamma(\bx_{i\perp},\bx_{j\perp}) \bigr] \notag\\
 &\qquad -\sum_{i,j=1}^4 {\tf}^a_i\,{\tf}^{a\ast}_{\bar{j}}\,
  \varGamma(\bx_{i\perp},\bx_{j\perp}) \biggr\} \,,
\end{align}
where we use the same notation as in Ref.~\cite{Fukushima:2007dy} as
\begin{equation}
 \begin{split}
 L(\bx_{1\perp},\bx_{2\perp}) &= g^4\int\rmd^2\zt\,
  G_0(\bx_{1\perp}\!-\!\zt)\,G_0(\bx_{2\perp}\!-\!\zt)\,,\\
 \varGamma(\bx_{1\perp},\bx_{2\perp}) &=
  2L(0,0)-2L(\bx_{1\perp},\bx_{2\perp}) \,.
 \end{split}
\end{equation}
Here $G_0(\xt\!-\!\zt)$ is the propagator in two dimensions satisfying
$\partial_\perp^2 G_0(\xt\!-\!\zt)=\delta^{(2)}(\xt-\zt)$.  We note
that the saturation scale $\Qs'$ has a different normalization from
Ref.~\cite{Fukushima:2007dy}, that is, we have defined
$\Qs'^2=2\Nc^2/(\Nc^2-1)\Qs^2$.  By means of the bra- and ket-vectors,
the initial and final states become
\begin{equation}
 \begin{split}
  |\text{initial}\rangle &= \frac{1}{2}\Nc^2\bigl(
   |4123\rangle +|2341\rangle -|3142\rangle -|2413\rangle
   \bigr) \,,\notag\\
  \langle\text{final}| &= \Nc^2\langle\text{4321}|
   +\langle\text{1234}| -\Nc\langle\text{1324}|
   -\Nc\langle\text{4231}| \,,
 \end{split}
\end{equation}
where we define
\begin{equation}
 \langle a_1,a_2,a_3,a_4,\bar{a}_1,\bar{a}_2,\bar{a}_3,\bar{a}_4
  |ijkl\rangle = \frac{1}{\Nc^2}\delta_{a_1\bar{a}_i}\delta_{a_2\bar{a}_j}
  \delta_{a_3\bar{a}_k}\delta_{a_4\bar{a}_l} \,,
\end{equation}
whose norm is normalized properly to unity.  Here we note that
$|ijkl\rangle$'s form a complete set of the singlet states out of
$\Nc\otimes\Nc^*\otimes\Nc\otimes\Nc^*\otimes\Nc\otimes\Nc^*
 \otimes\Nc\otimes\Nc^*$, that is, satisfy $H_0 |ijkl\rangle=0$,
though $|ijkl\rangle$ is not the orthogonal basis.

  Let us consider the large-$\Nc$ limit and then one might naively
think that the leading order of the matrix element would be
$\mathcal{O}(\Nc^4)$ since both the initial and final states are
$\mathcal{O}(\Nc^2)$.  This is, however, not true.  In the large-$\Nc$
limit $V$ is diagonal and
$\langle\text{final}|\text{initial}\rangle=\mathcal{O}(\Nc^3)$~\cite{Marquet:2007vb,Fukushima:2007dy}.
Unfortunately, since the leading-order contribution vanishes, we need
to calculate the next-to-leading order in the large $\Nc$ expansion.

  For the purpose of proceeding to the next-to-leading order, we need
to know the overlap between different states.  These are directly read
as
\begin{equation}
 \begin{split}
  \langle 4321|4123\rangle &= \langle 4321|2341\rangle
   = \frac{1}{\Nc} \,,\\
  \langle 4321|3142\rangle &= \langle 4321|2413\rangle
   = \frac{1}{\Nc^2} \,.
\label{eq:overlap}
 \end{split}
\end{equation}
As we mentioned above, the matrix elements of $\rme^{-V}$ are
$\mathcal{O}(1)$ for the (vanishing) diagonal components and  are
$\mathcal{O}(1/\Nc)$ for the off-diagonal components.  Hence the only
first equation in Eq.~(\ref{eq:overlap}) gives a finite contribution
to the next-to-leading order, while the second one is the
next-to-next-to-leading order in this counting.

  As a result of the truncation up to $\mathcal{O}(\Nc^3)$ the matrix
element is expressed as
\begin{equation}
 \langle\text{final}|\rme^{-V}|\text{initial}\rangle \simeq \frac{\Nc^4}{2}
  \bigl(\langle 4321|\rme^{-V}|4123 \rangle
  +\langle 4321|\rme^{-V}|2341 \rangle \bigr) \,.
\label{eq:matrixelement}
\end{equation}

  Henceforth, let us focus on the evaluation of
$\langle 4321|\rme^{-V}|4123 \rangle$ appearing in
Eq.~(\ref{eq:matrixelement}).  By definition $\rme^{-V}$ is expanded
as
\begin{equation}
 \rme^{-V}=\sum_{n=0}^\infty\frac{(-1)^n}{n!}V^n \,.
\end{equation}
We shall operate $V$ onto $|4123\rangle$ and $|4321\rangle$.  Using
the diagrammatical technique developed in Ref.~\cite{Fukushima:2007dy}
we can find,
\begin{equation}
 \begin{split}
  V|4123\rangle &= V_{11}|4123\rangle + \frac{1}{\Nc}
   V_{21}|4321\rangle + \cdots\,,\\
  V|4321\rangle &= V_{22}|4321\rangle + \cdots \,,
 \end{split}
\label{eq:Vact}
\end{equation}
where
\begin{align}
 V_{11} &= \frac{\Qs'^2}{2}\bigl[\varGamma(\bx_{1\perp},\bx_{2\perp})
  +\varGamma(\bx_{2\perp},\bx_{3\perp})+\varGamma(\bx_{3\perp},\bx_{4\perp})
  +\varGamma(\bx_{1\perp},\bx_{4\perp})\bigr]\,,\notag\\
 V_{21} &= \frac{\Qs'^2}{2}\bigl[\varGamma(\bx_{1\perp},\bx_{4\perp})
  +\varGamma(\bx_{2\perp},\bx_{3\perp})-\varGamma(\bx_{2\perp},\bx_{4\perp})
  -\varGamma(\bx_{1\perp},\bx_{3\perp})\bigr]\,,\notag\\
 V_{22} &= \frac{\Qs'^2}{2}\bigl[ 2\varGamma(\bx_{1\perp},\bx_{4\perp})
  +2\varGamma(\bx_{2\perp},\bx_{3\perp})\bigr] \,.
\end{align}
Here the ellipsis ``$\cdots$'' in Eq.~(\ref{eq:Vact}) represents
sub-leading states such as $|3142\rangle$, $|2413\rangle$, and so on,
which are higher order.  We can go on this procedure to make $V^n$ act
onto $|4123\rangle$, and then we get the following;
\begin{align}
 V^n|4123\rangle &= |4123\rangle V_{11}^n
  + \frac{1}{\Nc}|4321\rangle V_{21}(V_{11}^{n-1} + V_{22}V_{11}^{n-2}
  + \cdots + V_{22}^{n-1} ) + \cdots \notag\\
 &= |4123\rangle V_{11}^n +\frac{1}{\Nc}|4321\rangle V_{21}
  \frac{V_{11}^n-V_{22}^n}{V_{11}-V_{22}} + \cdots \,.
\end{align}
Consequently the necessary matrix element is acquired as
\begin{equation}
 \langle 4321|\rme^{-V}|4123 \rangle
  = \frac{1}{\Nc}\rme^{-V_{11}}+\frac{1}{\Nc}
  \frac{V_{21}}{V_{11}-V_{22}}(\rme^{-V_{11}}-\rme^{-V_{22}})
  +\mathcal{O}\bigl(\frac{1}{\Nc^2}\bigr) \,.
\end{equation}
We take the same path to evaluate $\langle 4321|
\rme^{-V}|2341\rangle$ only to find that
\begin{equation}
 \langle 4321|\rme^{-V}|2341\rangle
  = \langle 4321|\rme^{-V}|4123\rangle \,.
\end{equation}
Finally we reach the answer,
\begin{equation}
 \begin{split}
  & \bigl\langle \tr[U(\bx_{1\perp})\ta^a U^\dagger(\bx_{2\perp})
   U(\bx_{3\perp})\ta^a U^\dagger(\bx_{4\perp})] \bigr\rangle \\
  &= \langle\text{finial}|\rme^{-V}|\text{initial}\rangle
   =\Nc^3 \Bigl[\rme^{-V_{11}}+\frac{V_{21}}{V_{11}-V_{22}}
   (\rme^{-V_{11}}-\rme^{-V_{22}}) \Bigr] \,.
\end{split}
\end{equation}
Now that we have the four-point function, it is straightforward to
evaluate the two- and three-point functions by substituting
$(\bx_{2\perp}\!\to\bx_{3\perp}, \bx_{4\perp}\!\to\bx_{2\perp})$ and
$\bx_{4\perp}\!\to\bx_{3\perp}$, respectively.  That is,
\begin{equation}
 \Nc \bigl\langle\tr[U(\bx_{1\perp})U^\dagger(\bx_{2\perp})]\bigr\rangle
  = \Nc^3\,\rme^{-V_{22}} = \Nc^3\,
  \rme^{-\Qs'^2\varGamma(\bx_{1\perp},\bx_{2\perp})} \,,
\end{equation}
and
\begin{equation}
 \begin{split}
 & \bigl\langle\tr[U(\bx_{1\perp})\ta^a U^\dagger(\bx_{2\perp})
  \ta^b]U^{\dagger ab}(\bx_{3\perp})\bigr\rangle \\
 &= \Nc^3\,\rme^{-V_{22}} = \Nc^3\,
  \rme^{-\frac{1}{2}\Qs'^2[\varGamma(\bx_{1\perp},\bx_{2\perp})
  +\varGamma(\bx_{2\perp},\bx_{3\perp})+\varGamma(\bx_{1\perp},\bx_{2\perp})]} \,.
 \end{split}
\end{equation}
Although these results for the two- and three-point functions are
limited to large $\Nc$, we can reproduce the correct answer by
replacing $\Nc^3$ by $\Nc(\Nc^2-1)$ in view of the results available
from Refs.~\cite{Fukushima:2007dy,Kovner:2001vi}.


\end{document}